\documentclass[prd,preprint,showpacs,groupedaddress]{revtex4-1}
\usepackage{amsmath}
\pdfoutput=1
\usepackage{amsfonts}
\usepackage{amssymb}
\usepackage{geometry}
\usepackage{natbib}
\usepackage[english]{babel}
\usepackage{graphicx}
\usepackage{epstopdf}
\usepackage{subfigure}
\usepackage{caption}
\usepackage{multirow}
\usepackage{indentfirst}
\usepackage{mathrsfs}
\usepackage[colorlinks,linkcolor=blue,anchorcolor=blue,citecolor=blue]{hyperref}

\begin{document}

  \setlength{\parindent}{2em}
  \title{Optical properties of a nonlinear magnetic charged rotating black hole surrounded by quintessence with a cosmological constant }
  \author{Yuan Chen} \author{He-Xu Zhang} \author{Tian-Chi Ma} %\author{} 
  \author{Jian-Bo Deng} \email[Jian-Bo Deng: ]{dengjb@lzu.edu.cn}
  \affiliation{Institute of Theoretical Physics $\&$ Research Center of Gravitation, Lanzhou University, Lanzhou 730000, China}
  \date{\today}

  \begin{abstract}
  
  In this paper,we discuss optical properties of the nonlinear magnetic charged black hole surrounded by quintessence with a non-zero cosmological constant  $\Lambda$. Setting the state parameter $\omega=-3/2$ , we studied the horizon, the photon region and the shadow of this black hole. It turned out that for a fixed quintessential parameter $\gamma$, in a certain range, with the increase of the rotation parameter $a$ and magnetic charge $Q$, the inner horizon radius increases while the outer horizon radius decreases. And the cosmological horizon $r_{\Lambda}$ decrease when $\gamma$ or $\Lambda$ incease and increase slightly with increasing $a$ and $Q$. The shapes of photon region were then studied and depicted through graphical illustrations. Finally, we discussed the effects of the quintessential parameter $\gamma$ and the cosmological constant  $\Lambda$ on the shadow cast by this balck hole with a fixed observer position. 
  \end{abstract}

  \maketitle

\section{Introduction}
   Modern cosmological observations reveal that most galaxies may contain supermassive black holes in their centers. As an instance, astrophysical observations expect that there is a black hole lying at the center of our galaxy. Recently, a project, named the Event Horizon Telescope\cite{doeleman2008event}, is successfully collecting signals from radio sources. And it may help us  testing the theory of general relativity in the strong-field regime. So it is necessary to study the 
   optical properties theoretically, so that we can provide a reference substance  for future observation. 
   \par
    The photon sphere is a spherical surface round a black hole without rotation, which containing all the possible closed orbits of a photon. But for a balck hole with rotation, there is no longer a sphere but a region which is breaked from the photon sphere. It is a region which is filled by spherical lightlike geodesics. The light rays that asymptotically spiral towards one of these spherical lightlike geodesics corresponds to  boundary of the shadow. So the property of photon region takes great significance to the research of black hole shadow, which is an important project in black hole observation. 
    \par
    The shadow of a black hole is a natural consequence from the general theory of gravity. And even there is many the diﬀerent methods employed to determine the nature of the black hole, observation of the shadow of a black hole remains probably the most interesting one. The shadow of a spherically symmetric black hole appear as a perfect circle, which was first studied by Synge\cite{synge1966escape}. And then the spherically symmetric black hole with an accretion disk was disceussed by Luminet\cite{luminet1979image}. The shapes of shadows of black holes with ratation take deformation, and Bardeen was the first one to correctly calculate it\cite{novikov1973black}. Recent astrophysical observation have motivated many authors to study of shadows in theoretical investigation, like Kerr-Newman black holes\cite{de2000apparent}, naked singularities with deformation parameters\cite{hioki2009measurement},  multi-black holes\cite{yumoto2012shadows} and Kerr-NUT spacetimes \cite{abdujabbarov2013shadow}. And some authors tried to test theories of gravity by using the observations of shadow obtained from the black hole lying in our galaxy\cite{bambi2009apparent,johannsen2016testing2,broderick2014testing}.
\par
  In present view, the cosmological constant has been considered to have the potential to explain several theoretical and observational problems, for example, as a possible cause of the observed acceleration of the universe. And one of the strongest supports was the results of the observation for a supernova\cite{riess1998observational}, which suggests the probable presence of a positive cosmological constant for our universe. Therefore studying black holes with a non-zero cosmological constant has an significant place in research. Many authors studied black holes with a non-zero cosmological constants\cite{gibbons2004rotating,bakopoulos2019novel,wei2020extended,perlick2018black,firouzjaee2019black} And it should be noted that in the case with a non-zero cosmological constant the traditional method to compute shadow, which place observers at near infinite distance, is no longer valid. So we will take a different method, following\cite{grenzebach2014photon}.
\par
  Recent cosmological observations suggest that our current universe contains mainly of 68.3$\%$ dark energy , 26.8$\%$ dark matter, and 4.9$\%$ baryon matter, according to the Standard Model of Cosmology \cite{ade2016planck}. Thus, it is necessary to consider dark matter or dark energy in the black hole solutions. In recent years, the black hole surrounded by quintessence dark energy have caught a lot of attention. For example, Kiselev \cite{kiselev2003quintessence} considered the Schwarzschild black hole surrounded by the quintessential energy and then Toshmatov and Stuchl\'ik \cite{toshmatov2017rotating} extended it to the Kerr-like black hole; the quasinormal modes, thermodynamics and phase transition from Bardeen Black hole surrounded by quintessence was discussed by Saleh and Thomas \cite{saleh2018thermodynamics}; the Hayward black holes surrounded by quintessence have been studied in Ref.~\cite{benavides2020rotating}, etc \cite{ghosh2018lovelock,zhang2006quasinormal,ghosh2016rotating,chen2008hawking,abdujabbarov2017shadow,azreg2013thermodynamical}, see Refs.~\cite{haroon2019shadow,tsujikawa2013quintessence,zhang2020optical,ghaffarnejad2018quintessence,sakti2019kerr,ghosh2017quintessence,ma2017thermodynamic,zhang2020bardeen} for more recent research. And some authors have considered black holes surrounded by quintessence with a non-zero cosmological constant\cite{hong2019thermodynamics,xu2017kerr,chabab2018more,azreg2015charged,chen2013holographic}. And for discussing the effects of cosmological background of black holes, we will consider quintessence in our work. 
\par
  Our paper is organized as follows. In the next section, we give the solution of nonlinear magnetic charged rotating black hole surrounded by quintessence including a cosmological constant. The horizons is the subject of Sec.~\ref{4}. In Sec.~\ref{5}, we study the photon region. And the shadows are discussed in Sec.~\ref{6}. Conclusions and discussions are presented in Sec.~\ref{7}. For simplicity, the quintessential state parameter $\omega$ and mass $M$ will be set to -2/3 and 1, respectively.

\section{the metric of nonlinear magnetic charged rotating black hole surrounded by quintessence with a cosmological constant}\label{II}
  Kiselev first derived the solutions of the black hole surrounded by the quintessence \cite{kiselev2003quintessence}. Then, it have been applied in many articles. As a result of applying, the solution of nonlinear magnetic-charged black hole surrounded by quintessence was obtained in Ref.\cite{nam2018non} by considering Einstein gravity coupled to a nonlinear electromagnetic field.
  And recently the solution has been generalized to include rotation\cite{benavides2020rotating}. And the rotational version is given by
 \begin{equation}\label{eq:rotating metric 1}
  \mathrm{d}s^{2}=-\frac{\Delta_r}{\Sigma}\left(\mathrm{d}t-a\sin^{2}{\theta}\mathrm{d}\phi \right)^{2}+\frac{\Sigma}{\Delta_{r}}\mathrm{d}r^{2}+\Sigma\mathrm{d}\theta^{2}+\frac{\sin^{2}{\theta} }{\Sigma}\left(a\mathrm{d}t-\left(r^{2}+a^{2}\right)\mathrm{d}\phi \right)^{2}
  \end{equation}
  with
  \begin{equation}\label{eq:line element terms}
  \begin{gathered}
  \Sigma=r^{2}+a^{2}\cos^{2}{\theta}, \Delta_{r}=r^{2}+a^{2}-\frac{2Mr^{4}}{r^{3}+Q^{3}}-\gamma r^{1-3\omega}.
  \end{gathered}
  \end{equation}
  where the $a$ is a rotional parameter, $Q$ is magnetic charge, $\omega$ is a state parameter and $\gamma$ is the quintessential paramater.
  \par
 Considering Einstein equation with a positive cosmological constant, it takes the form 
   \begin{equation}\label{eq:Ein}
  \begin{gathered}
   \widetilde{G}_{\mu\nu}=R_{\mu\nu}-\frac{1}{2} Rg_{\mu\nu}+\Lambda g_{\mu\nu}=8\pi T_{\mu\nu}.
  \end{gathered}
  \end{equation}

  With the cosmological constant, we assume the solution including a cosmological constant is
  \begin{equation}\label{eq:rotating metric}
  \begin{split}
  \mathrm{d}s^{2}=&-\frac{\Delta_r}{\Xi^{2}\Sigma}\left(\mathrm{d}t-a\sin^{2}{\theta}\mathrm{d}\phi \right)^{2}+\frac{\Sigma}{\Delta_{r}}\mathrm{d}r^{2}+\frac{\Sigma}{\Delta_\theta}\mathrm{d}\theta^{2}+\frac{\Delta_\theta\sin^{2}{\theta} }{\Xi^{2}\Sigma}\left(a\mathrm{d}t-\left(r^{2}+a^{2}\right)\mathrm{d}\phi \right)^{2},
  \end{split}
  \end{equation}
  with
  \begin{equation}
  \begin{gathered}
  \Xi = 1+\frac{\Lambda}{3}a^{2},\Sigma=r^{2}+a^{2}\cos^{2}{\theta},
  \Delta_{\theta}=1+\frac{\Lambda}{3}a^{2}\cos^{2}{\theta},\\
  \Delta_{r}=r^{2}+a^{2}-\frac{2Mr^{4}}{r^{3}+Q^{3}}-\gamma r^{1-3\omega}-\frac{\Lambda}{3}\left(r^{2}+a^{2}\right)r^{2}.
  \end{gathered}
 \end{equation}
\par
  With the help of Mathmatica, we find the metric (\ref{eq:rotating metric}) accutually satisfies Einstein equation (\ref{eq:Ein}) coupling with a non-linear electromagnetic field in quintessence with a cosmological constant. The solution reduce to nonlinear magnetic charged solution when $\Lambda=0$, $a=0$, and $\gamma=0$. And when $\Lambda=0$, $a=0$, $Q=0$ and $\gamma=0$, it come back to Schwarzschild spacetime.

\section{Horizon}\label{4}
  Similar to the Kerr black hole, the space-time metric (\ref{eq:rotating metric}) is singular at $\Delta_{r}=0$, which corresponds to the horizons of the rotating black hole. In other words, the horizons are solutions of
  \begin{equation}\label{eq:horziom}
   \Delta_{r}=r^{2}+a^{2}-\frac{2Mr^{4}}{r^{3}+Q^{3}}-\gamma r^{3}-\frac{\Lambda}{3}\left(r^{2}+a^{2}\right)r^{2}=0.
  \end{equation}
\par
\begin{figure}[htbp]
  	\centering
  	\subfigure{
  		\includegraphics[width=.45\textwidth]{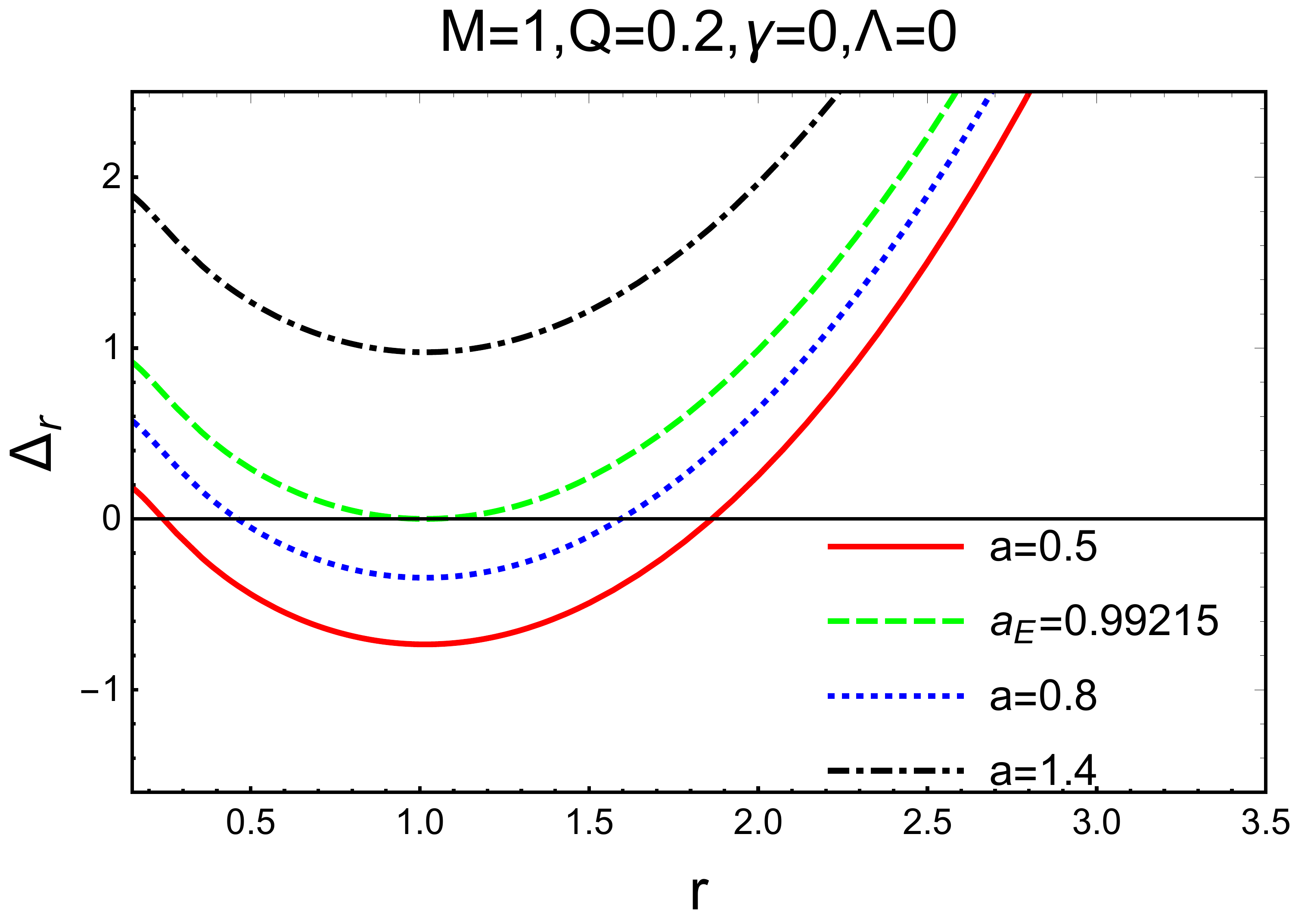}
  		\includegraphics[width=.45\textwidth]{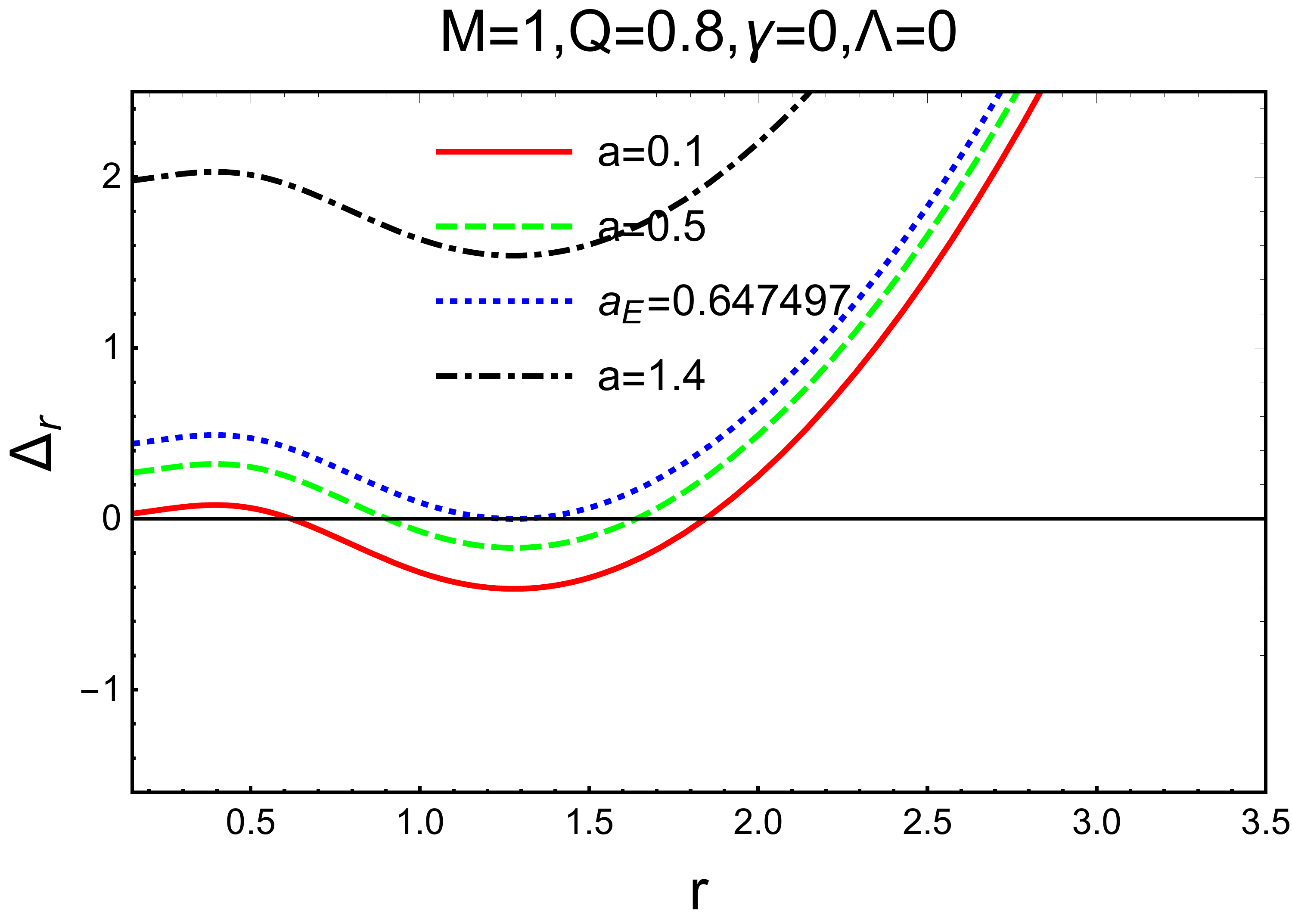}
  	}
  	\subfigure{
  		\includegraphics[width=.45\textwidth]{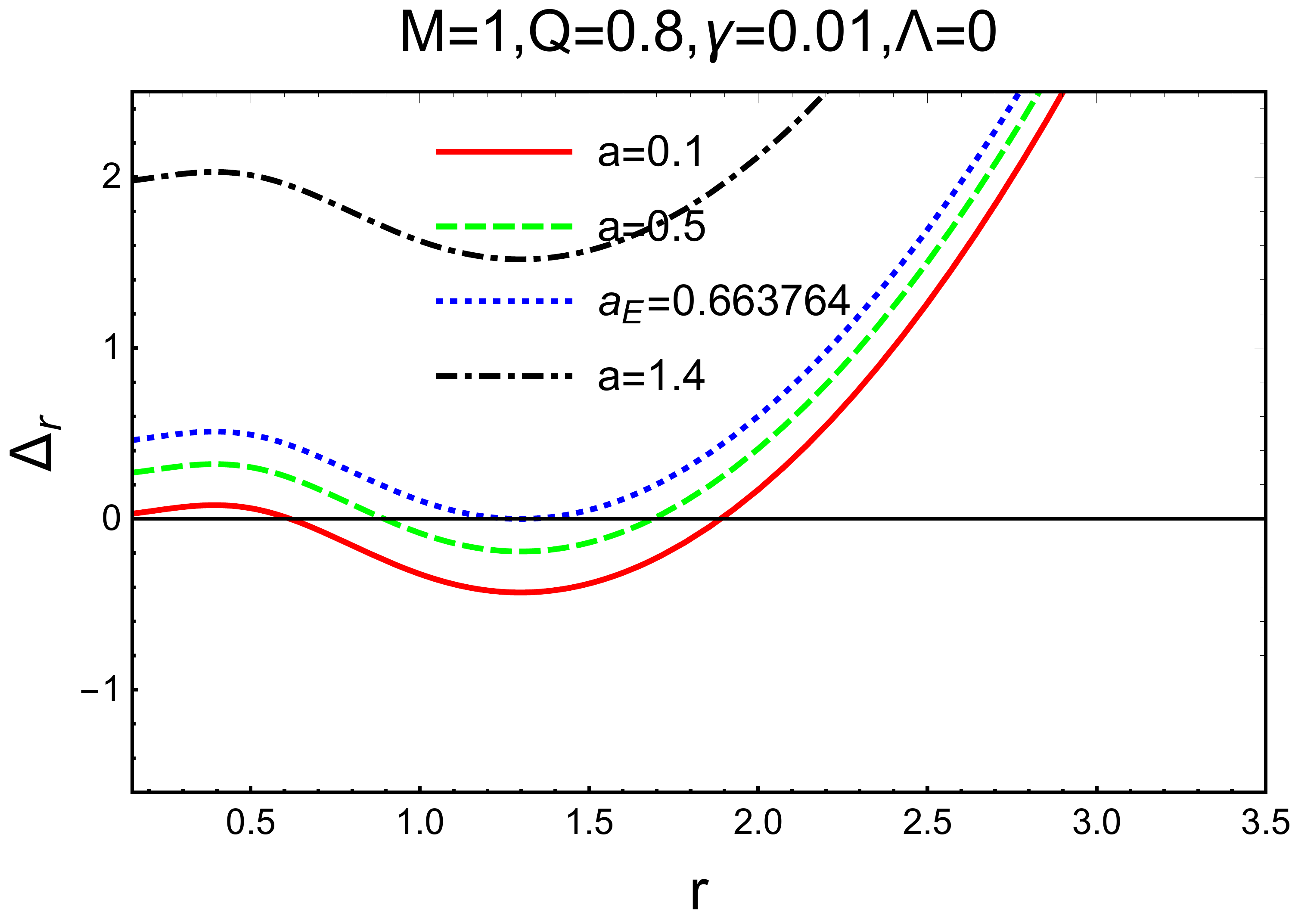}
  		\includegraphics[width=.45\textwidth]{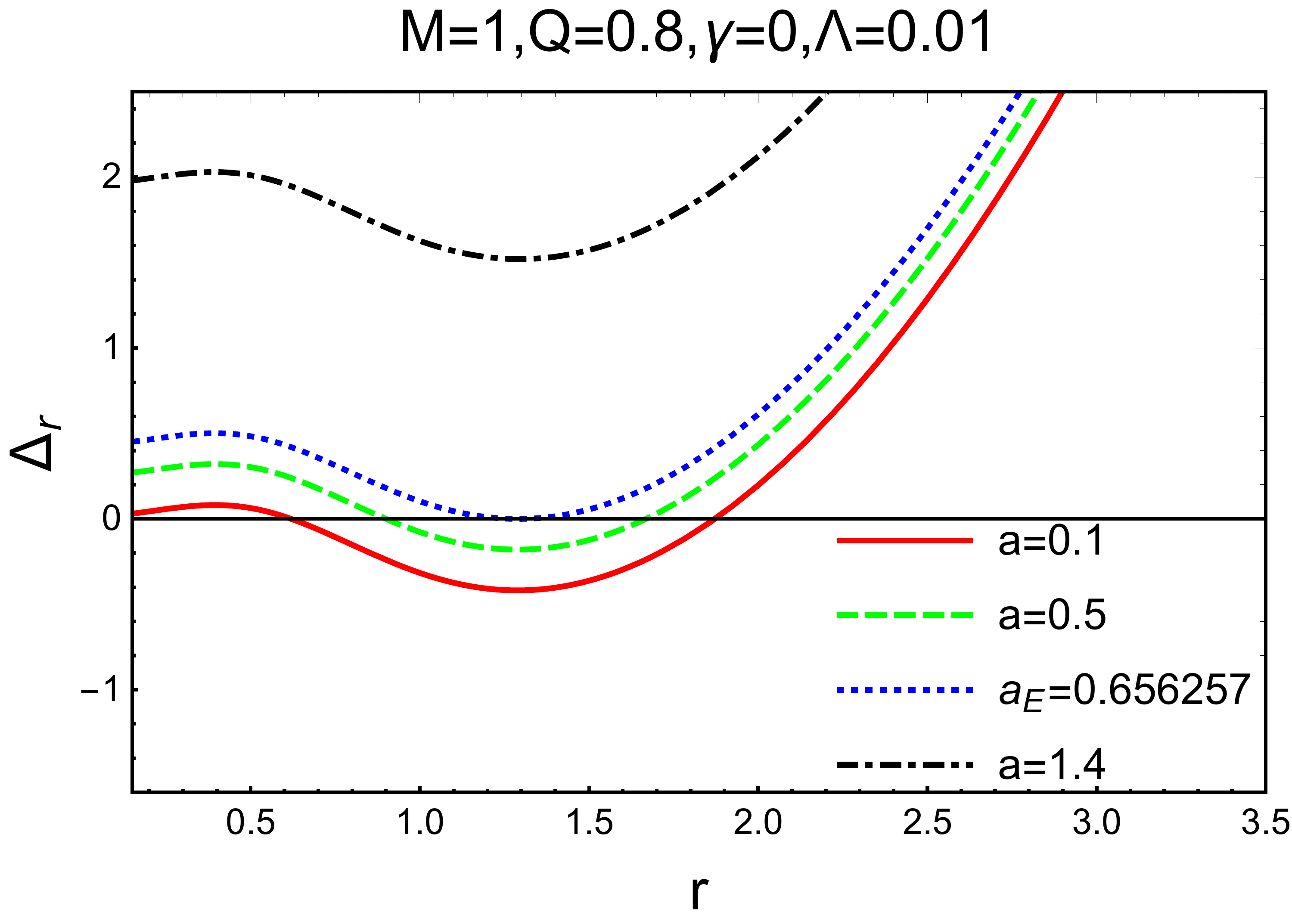}
  	}
  	\caption{Plot showing the behavior of horizons vs $r$ for a set of fixed values of $Q$, $\gamma$ and $\Lambda$ by varying $a$.}
  	\label{fig:horizon1}
  \end{figure}

  Obviously, the radii of horizons depend on the rotation parameter $a$, magnetic charge $Q$, quintessence parameter $\gamma$ and the cosmological constant $\Lambda$. This equation is unable to solved analytically. When $\Lambda$ is positive, the numerical analysis of Eq.~(\ref{eq:horziom}) suggests the possibility of three roots for a set of values of parameters, which correspond the inner horizon $r_{-}$ (smaller root), the outer horizon $r_{+}$ (larger root) and  the cosmological horizon $r_{\Lambda}$(the largest root), respectively. The variation of $\Delta_{r}$ with respect to $r$ for the different values of parameters $a$, with fixed $Q$, $\gamma$ and $\Lambda$  is depicted in Figs.~\ref{fig:horizon1} and the roots of $\Delta_{r}=0$ with varying $a$ has been listed in Tab.~\ref{tab:horizon1} and Tab~\ref{tab:horizon2}. As can be seen from Fig.~\ref{fig:horizon1}, Tab.~\ref{tab:horizon1} and Tab~\ref{tab:horizon2}, for fixed parameters $Q$, $\gamma$ and $\Lambda$, if $a<a_{E}$, the radii of outer horizons decrease with the increasing $a$ while the radii of inner horizons increase. For $a=a_{E}$, $r_{-}$ and $r_{+}$ meet at $r_{E}$, i.e. we have an extreme case which inner horizons and outer horizons degenerate. The critical rotation parameter $a_{E}$ and the corresponding critical radius $r_{E}$ can be calculated by combining $\Delta_{r}=0$ with $\frac{\mathrm{d}\Delta_{r}}{\mathrm{d}r}=0$. When $a>a_{E}$, inner horizon and outer horizon will not exists any more, which means there is no longer a black hole. At the same time, $r_{\Lambda}$ increase as $a$ increase but not so obviously.  
  \par
  
 \begin{figure}[bp]
  	\centering
  	\subfigure{
  		\includegraphics[width=.45\textwidth]{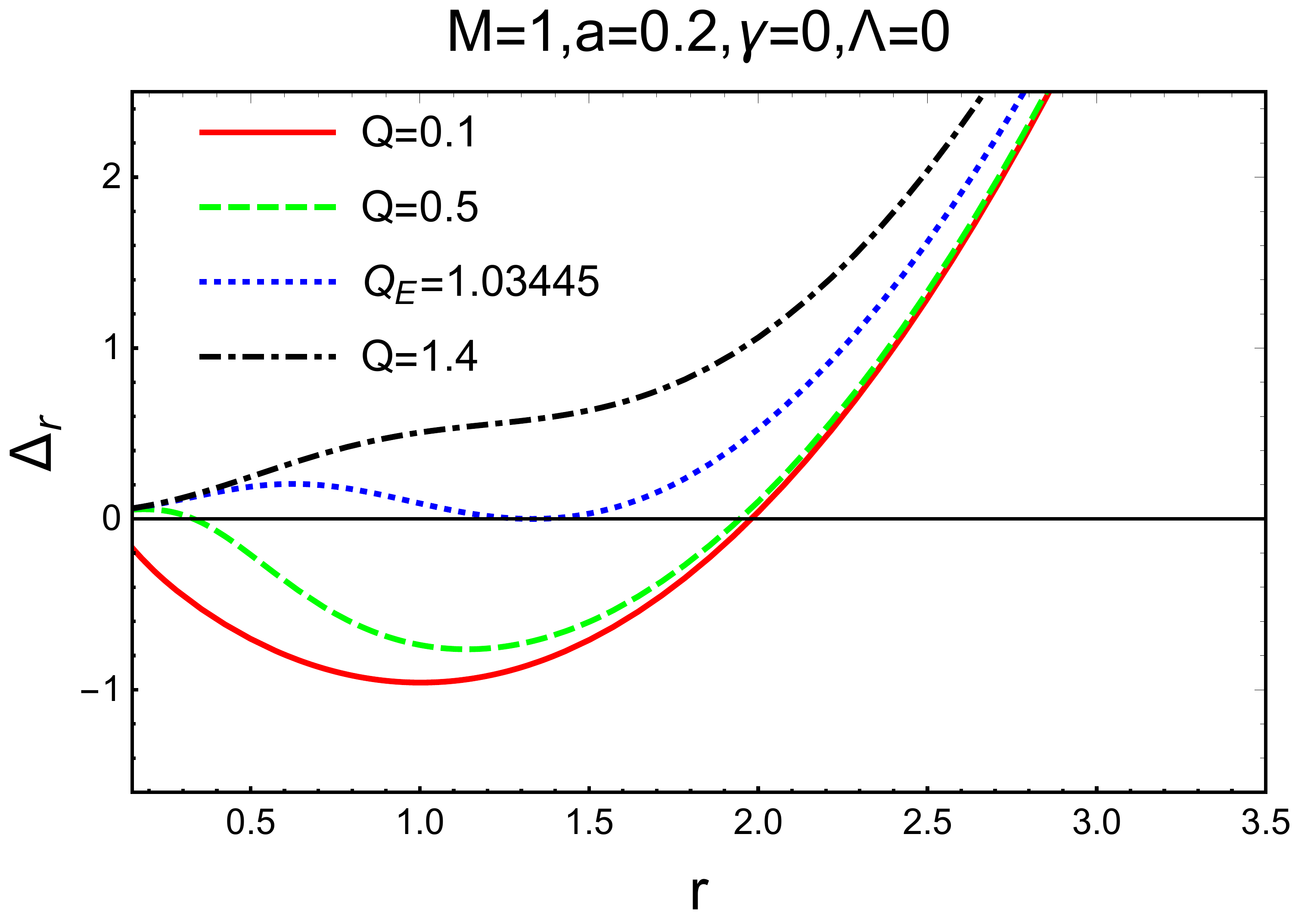}
  		\includegraphics[width=.45\textwidth]{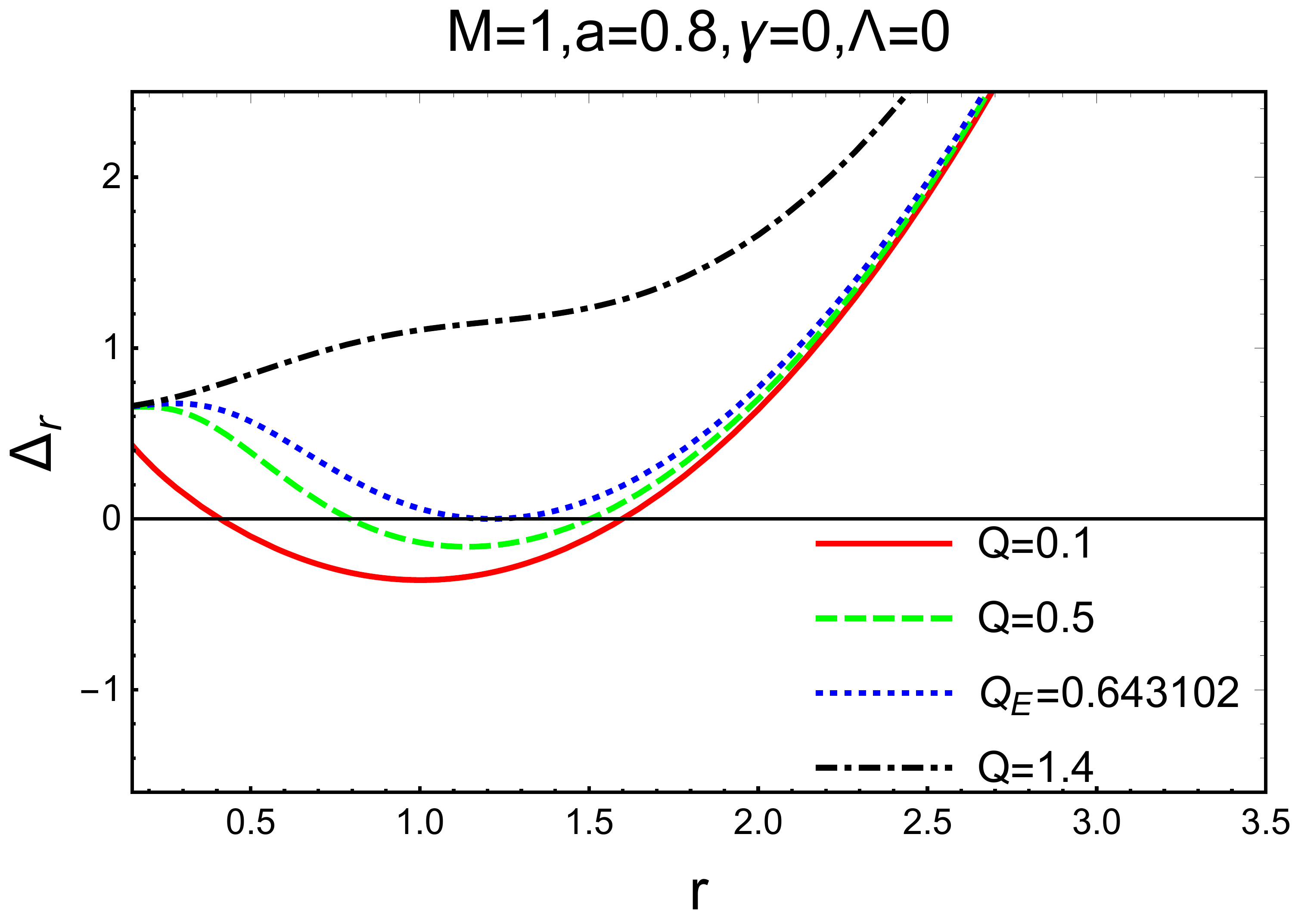}
  	}
  	\subfigure{
  		\includegraphics[width=.45\textwidth]{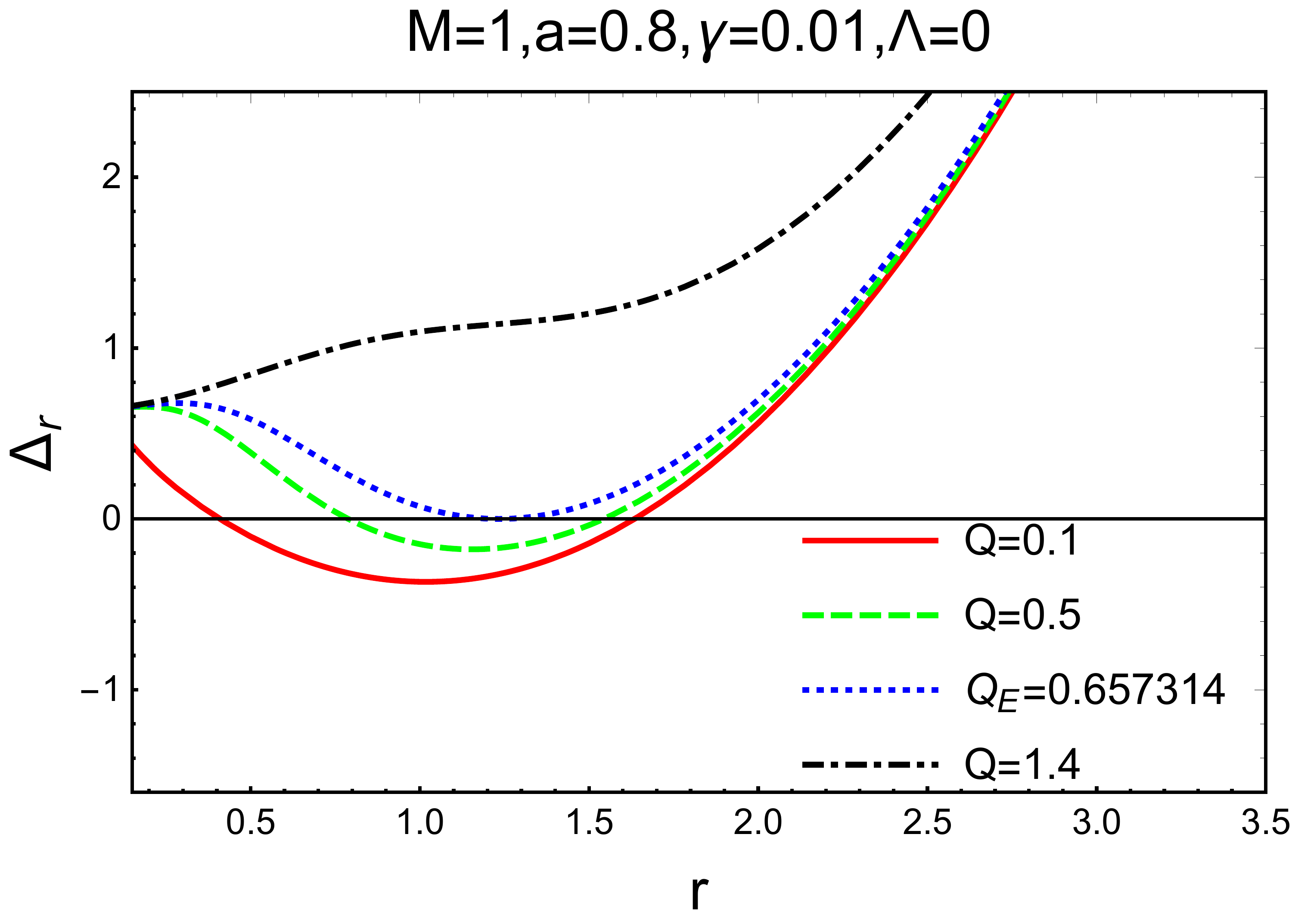}
  		\includegraphics[width=.45\textwidth]{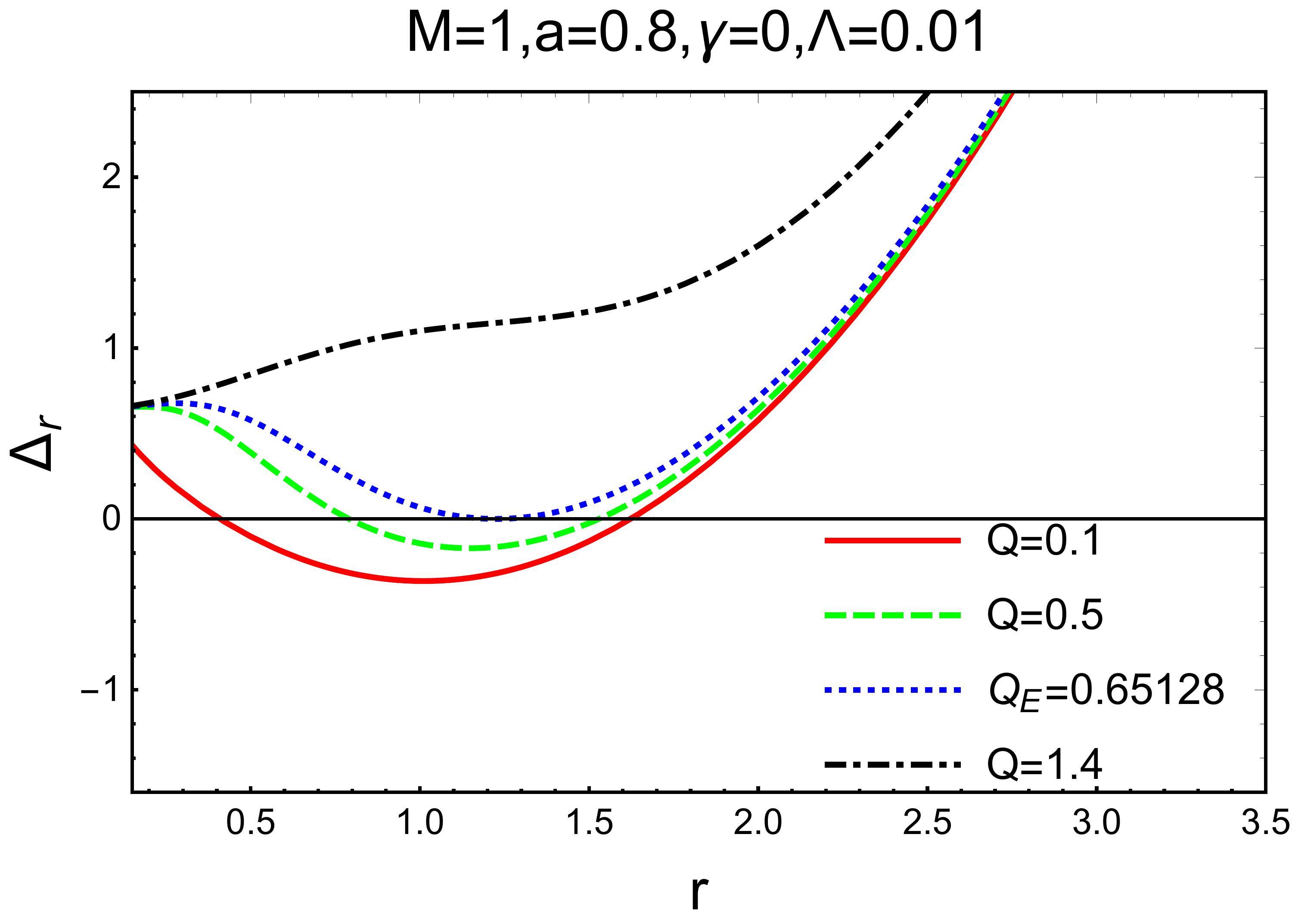}
  	}
  	\caption{Plot showing the behavior of horizons vs $r$ for a set of fixed values of $a$, $\gamma$ and $\Lambda$ by varying $Q$.}
  	\label{fig:horizon2}
  \end{figure}
A similar analysis can be applied to Q. We list the roots in Tab.\ref{tab:horizonq} with varying $Q$ in the case $a=0.2$, $\gamma=0.01$ and $\Lambda=0.02$ and depict the variation of $\Delta_{r}$ with respect to $r$ for the different values of parameters $Q$ with fixed $a$, $\gamma$ and $\Lambda$ in Fig.~\ref{fig:horizon2}. The result shows that, for any given values of parameters $a$, $\gamma$ and $\Lambda$, inner and outer horizons get closer first with the increase of $Q$, then coincide when $Q=Q_{E}$ and disappear when $Q>Q_{E}$. And $r_{\Lambda}$ increase as $Q$ increase but not so significantly.
\par
\renewcommand\arraystretch{0.6}  
\begin{table}[tb]
	\caption{The radii of horizons for a set of fixed values of $Q=0.2$ and $\Lambda=0.01$ by varying $a$ in different values of $\gamma$}
	\label{tab:horizon1}
	\begin{tabular}{l|l|l|l|l|l|l|l|l|l}
		\hline\hline
		a    &\multicolumn{3}{l|}{$\gamma=0.01$} & \multicolumn{3}{l|}{$\gamma=0.02$} & \multicolumn{3}{l}{$\gamma=0.03$} \\ \hline
		&${r_{-}}$       &${r_{+}}$     &${r_{\Lambda}}$       &${r_{-}}$       &${r_{+}}$   &${r_{\Lambda}}$      &${r_{-}}$        &${r_{+}}$      &${r_{\Lambda}}$     \\   \hline
		0&    0.06427&    2.071  & 14.666   & 0.06424  & 2.120  & 13.2381  & 0.06422  & 2.175  & 11.930   \\ \hline
		0.1&  0.09616&    2.065  & 14.6661  & 0.09614  & 2.114  & 13.2384  & 0.09613  & 2.169  & 11.9308  \\ \hline
		0.2&  0.13072&    2.050  & 14.6665  & 0.13069  & 2.098  & 13.2391  & 0.13068  & 2.152  & 11.932   \\ \hline
		0.3&  0.16415&    2.023  & 14.6672  & 0.16413  & 2.071  & 13.2403  & 0.16410  & 2.124  & 11.9339  \\ \hline
		0.4&  0.19995&    1.984  & 14.6681  & 0.1999   & 2.031  & 13.242   & 0.19987  & 2.083  & 11.9366  \\ \hline
		0.5&  0.24154&    1.931  & 14.6693  & 0.24147  & 1.977  & 13.2442  & 0.24141  & 2.028  & 11.9401  \\ \hline
		0.6&  0.29343&    1.863  & 14.6708  & 0.29331  & 1.907  & 13.2469  & 0.29318  & 1.956  & 11.9443  \\ \hline
		0.7&  0.36245&    1.774  & 14.6725  & 0.36217  & 1.817  & 13.2501  & 0.36190  & 1.863  & 11.9492  \\ \hline
		0.8&  0.45992&    1.656  & 14.6745  & 0.45921  & 1.697  & 13.2537  & 0.45851  & 1.742  & 11.9549  \\ \hline\hline
		
	\end{tabular}
\end{table}

\begin{table}[tb]
	\caption{The radii of horizons for a set of fixed values of $Q=0.2$ and $\gamma=0.01$ by varying $a$ in different values of $\Lambda$.}
	\label{tab:horizon2}
	\begin{tabular}{l|l|l|l|l|l|l|l|l|l}
		\hline\hline
		a   & \multicolumn{3}{l|}{$\Lambda=0.01$}& \multicolumn{3}{l|}{$\Lambda=0.02$} & \multicolumn{3}{l}{$\Lambda=0.03$} \\ \hline
		&${r_{-}}$       &${r_{+}}$    &${r_{\Lambda}}$       &${r_{-}}$       &${r_{+}}$   &${r_{\Lambda}}$      &${r_{-}}$        &${r_{+}}$     &${r_{\Lambda}}$       \\ \hline
		0   & 0.06427  & 2.071  & 14.666   & 0.06426  & 2.105  & 10.2654  & 0.06426  & 2.143  & 8.2118  \\ \hline
		0.1 & 0.09616  & 2.065  & 14.6661  & 0.09616  & 2.099  & 10.2656  & 0.09616  & 2.137  & 8.2121  \\ \hline
		0.2 & 0.13072  & 2.050  & 14.6665  & 0.13072  & 2.083  & 10.2662  & 0.13071  & 2.121  & 8.2131  \\ \hline
		0.3 & 0.16415  & 2.023  & 14.6672  & 0.16414  & 2.056  & 10.2673  & 0.16414  & 2.093  & 8.2148  \\ \hline
		0.4 & 0.19995  & 1.984  & 14.6681  & 0.19993  & 2.016  & 10.2689  & 0.19992  & 2.052  & 8.2171  \\ \hline
		0.5 & 0.24154  & 1.931  & 14.6693  & 0.24151  & 1.963  & 10.2709  & 0.24148  & 1.997  & 8.2201  \\ \hline
		0.6 & 0.29343  & 1.863  & 14.6708  & 0.29337  & 1.893  & 10.2733  & 0.29330  & 1.926  & 8.2237  \\ \hline
		0.7 & 0.36245  & 1.774  & 14.6725  & 0.36229  & 1.803  & 10.2761  & 0.36213  & 1.835  & 8.2279  \\ \hline
		0.8 & 0.45992  & 1.656  & 14.6745  & 0.45948  & 1.684  & 10.2793  & 0.45905  & 1.714  & 8.2328  \\ \hline\hline
		
	\end{tabular}
\end{table}

\begin{table}[tb]
	\caption{The radii of horizons for a set of fixed values of $a=0.2$ and $\gamma=0.01,\Lambda=0.01$ by varying $Q$.}
	\label{tab:horizonq}
	\begin{tabular}{|l|l|l|l|}
		\hline\hline
		Q   & \multicolumn{3}{l|}{$\Lambda=0.01,\gamma=0.01$} \\ \hline
		&${r_{-}}$       &${r_{+}}$    &${r_{\Lambda}}$    \\ \hline
		0   & 0.02020  & 2.0853  & 10.2662    \\ \hline
		0.1 & 0.07565  & 2.0851  & 10.2662    \\ \hline
		0.2 & 0.11461  & 2.0832  & 10.2662    \\ \hline
		0.3 & 0.13071  & 2.0783  & 10.2663    \\ \hline
		0.4 & 0.18902  & 2.0687  & 10.2663    \\ \hline
		0.5 & 0.25485  & 2.0523  & 10.2664    \\ \hline
		0.6 & 0.33126  & 2.0268  & 10.2665    \\ \hline
		0.7 & 0.42101  & 1.9889  & 10.2667    \\ \hline
		0.8 & 0.65584  & 1.9329  & 10.267     \\ \hline\hline
		
	\end{tabular}
\end{table}
Then, we further analyze the behavior of the horizons under different values of quintessence paramater and cosmological constant. For simplicity , we set $Q=0.2$ and consider the different values of cosmological constant $\Lambda$ and quintessential parameter $\gamma$, then vary the value of $a$ in the interval $0<a<1$, find the roots and list them in Tab~\ref{tab:horizon1} and Tab~\ref{tab:horizon1}. From the Tables, we can see that with fixed $a$ the cosmological horizon $r_{\Lambda}$ significantly decrease while $r_{+}$ increase as $\gamma$ or $\Lambda$ increase.

\section{photon region}\label{5}
 In this section, we turn to the property of the photon region. For the spacetime (\ref{eq:rotating metric}), geodesic motion can be governed by the Hamilton Jacobi equation \cite{saleh2018thermodynamics},
\begin{equation}\label{eq:ha}
  \frac{\partial S}{\partial \tau}=-\frac{1}{2} g^{\mu\nu}\frac{\partial{S}}{\partial{x^{\mu}}}\frac{\partial{S}}{\partial{x^{\nu}}}.
  \end{equation}
 where the $\tau$ is the affine parameter, $x^{\mu}$ represents the 4-vector, and $S$ is the Hamilton-Jacobi action function for a test particle, which can be writed in following form:
\begin{equation}\label{eq:hj}
  S=\frac{1}{2} m^{2}\tau -Et + L\phi + S_{r}(r) + S_{\theta}(\theta).
  \end{equation}
\par
  Then, by inserting the metric tensor of spacetime (\ref{eq:rotating metric}) and Eq. (\ref{eq:hj}) to Eq. (\ref{eq:ha}), we obtain 
\begin{equation}\label{eq:sq}
\begin{split}  
m^{2}=\frac{\Xi^{2}((r^{2}+a^{2})^{2}\Delta_{\theta}-a^{2}\sin^{2}(\theta)\Delta_{r})}{\Delta_{\theta}\Delta_{r}\Sigma} E^{2}-\frac{\Delta_{r}}{\Sigma}(\frac{\partial{S_{r}(r)}}{\partial{r}})^{2} - \frac{\Delta_{\theta}}{\Sigma}(\frac{\partial{S_{\theta}(\theta)}}{\partial{\theta}})^{2}\\ -\frac{\Xi^{2}(-a^{2}\Delta_{\theta}+\csc^{2}(\theta)\Delta_{r})}{\Delta_{\theta}\Delta_{r}\Sigma} L^{2} -\frac{2a\Xi^{2}((r^{2}+a^{2})\Delta_{\theta}-\Delta_{r})}{\Delta_{\theta}\Delta_{r}\Sigma} EL.
 \end{split}
  \end{equation}
\begin{figure}[htp]
	\centering
	\subfigure{
		\includegraphics[width=.34\textwidth]{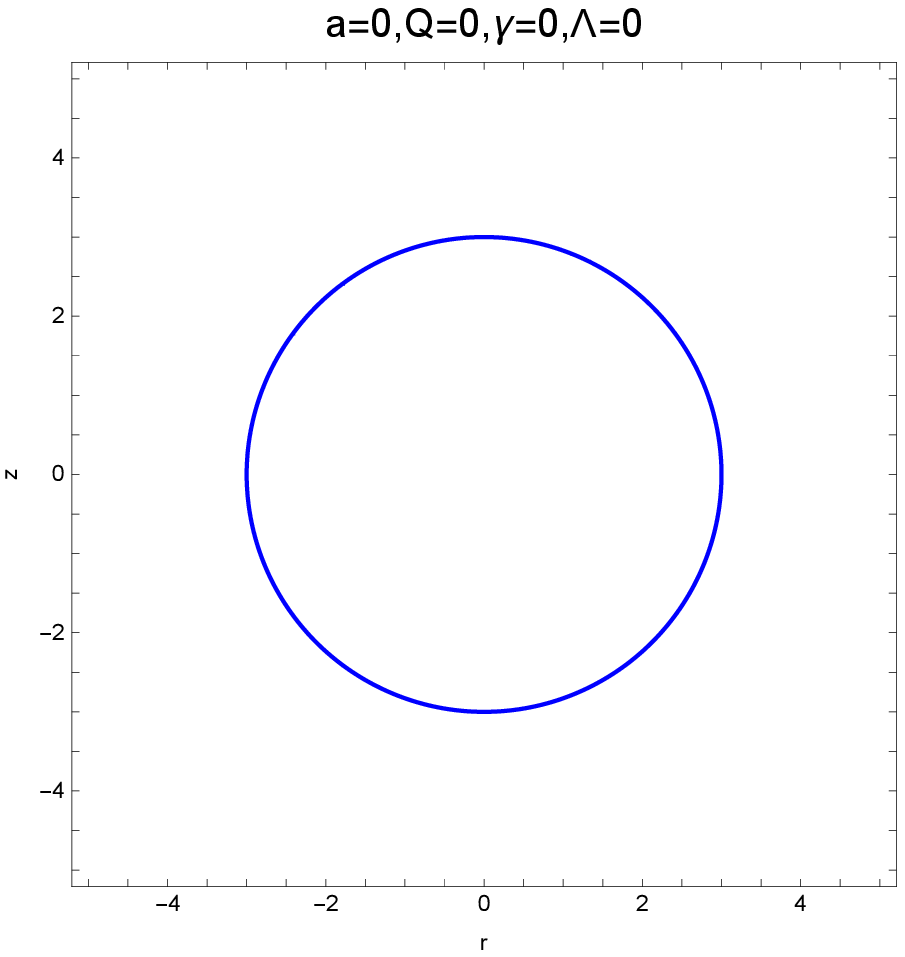}
		\includegraphics[width=.34\textwidth]{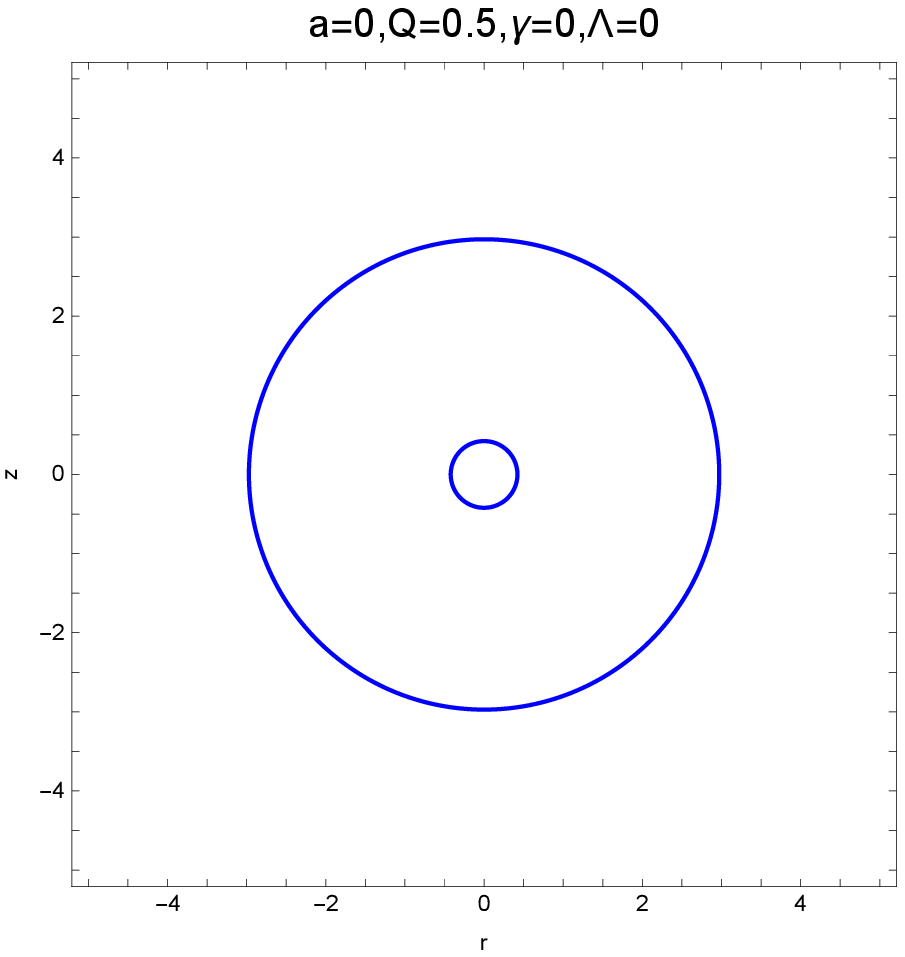}
		\includegraphics[width=.34\textwidth]{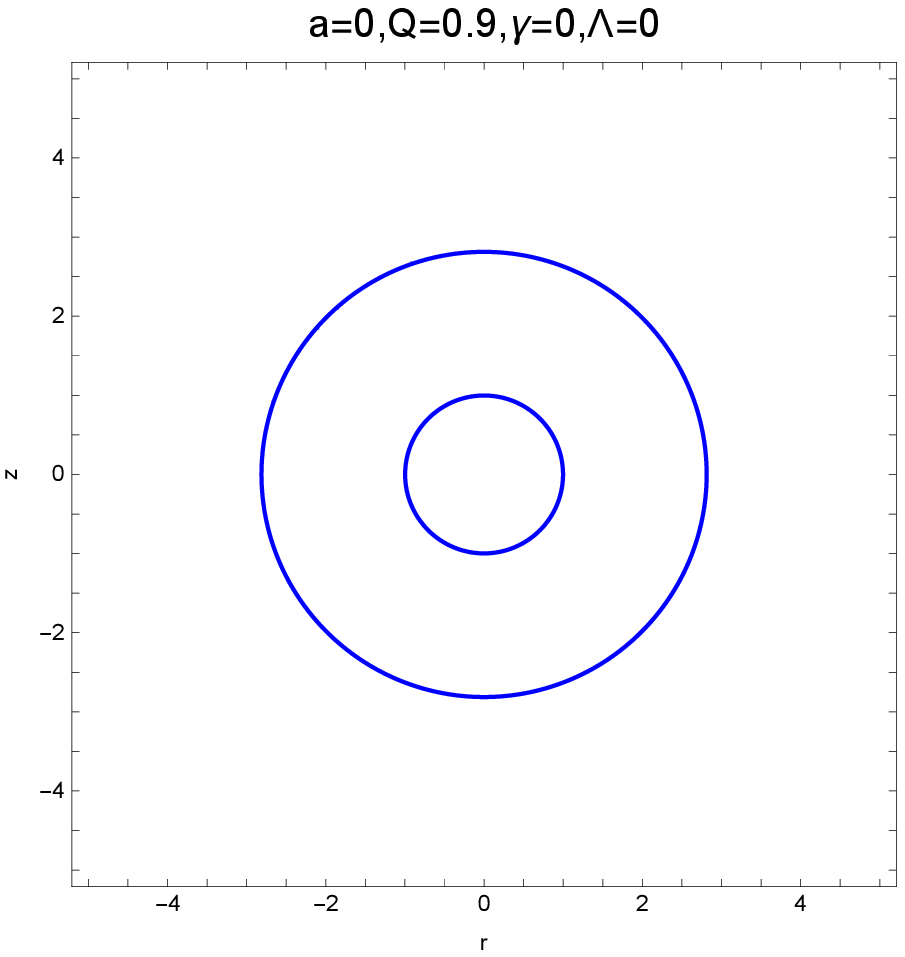}
	}
	\centering
	\subfigure{
		\includegraphics[width=.34\textwidth]{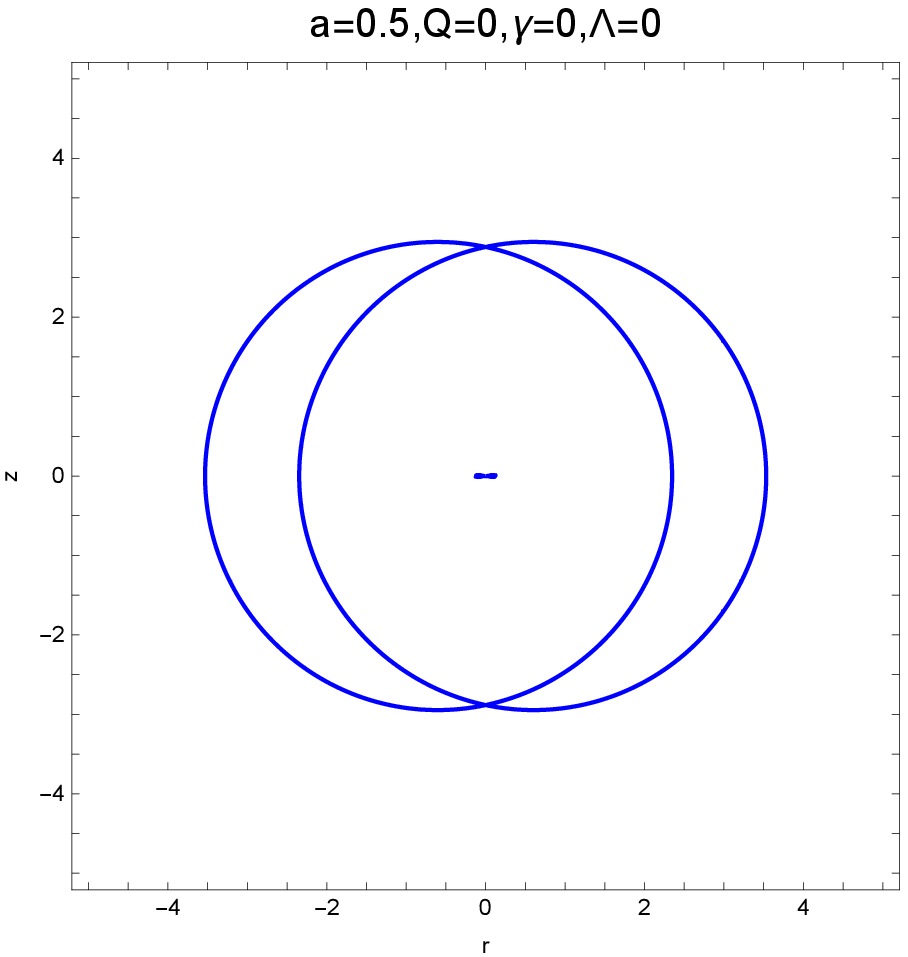}
		\includegraphics[width=.34\textwidth]{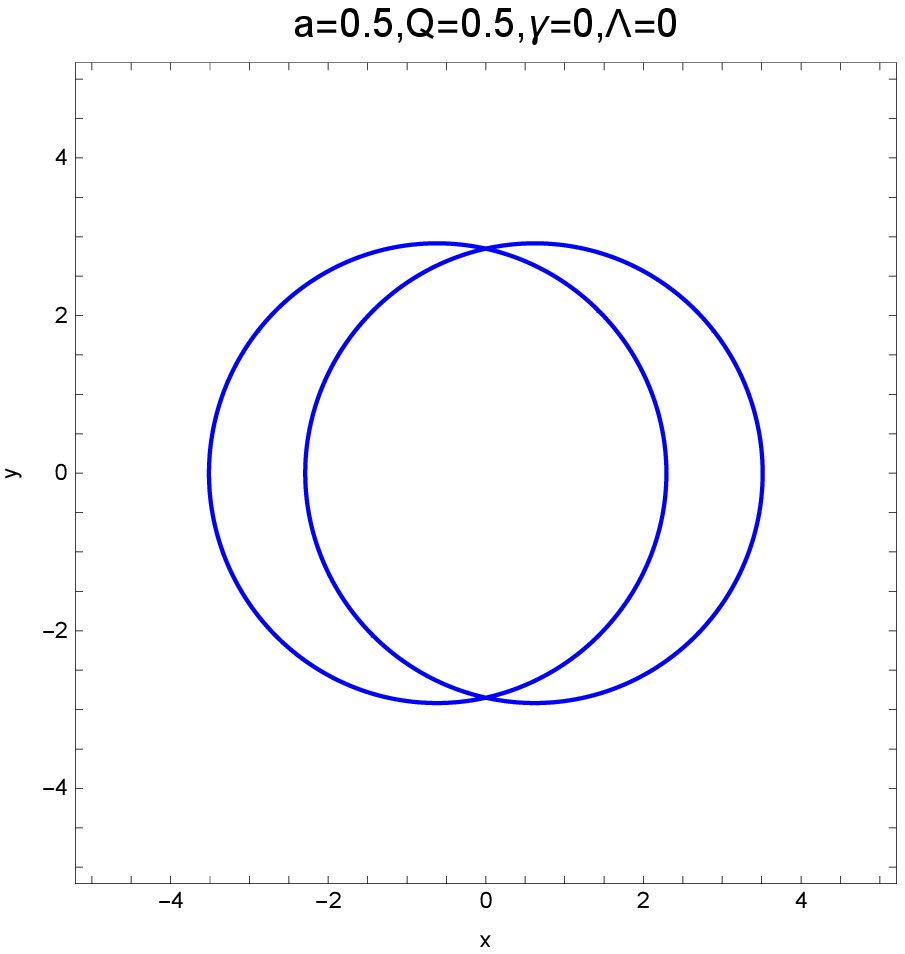}
		\includegraphics[width=.34\textwidth]{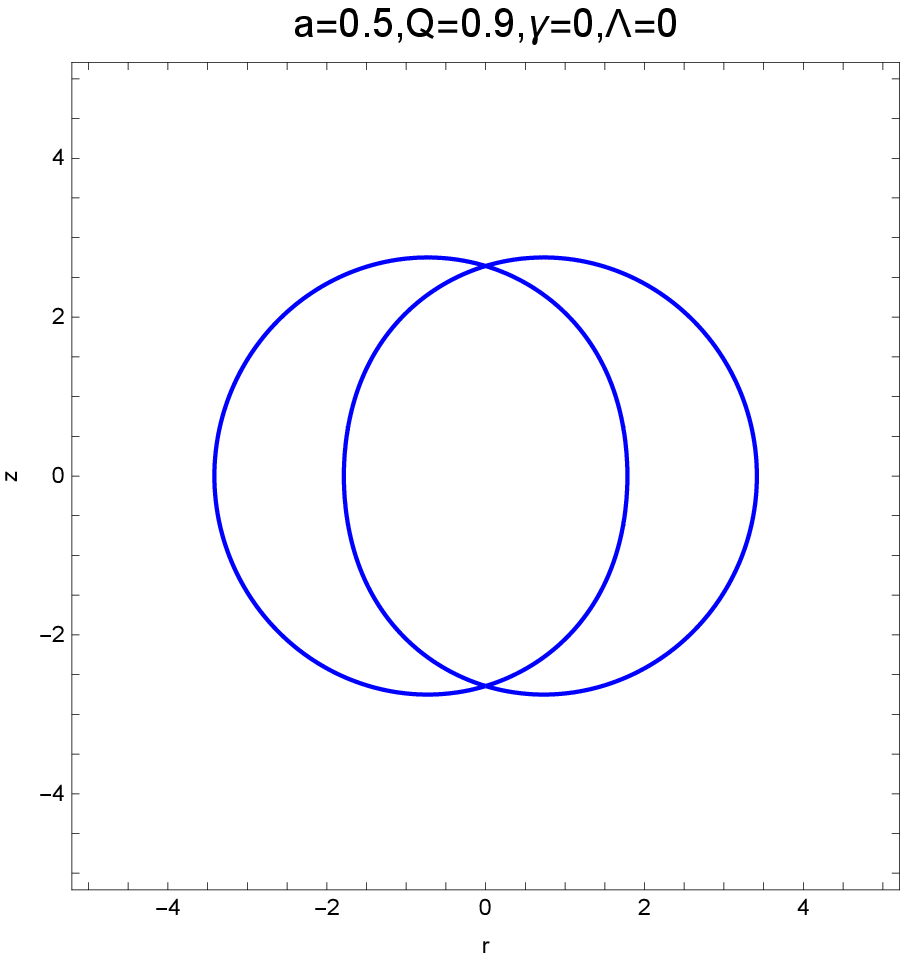}
	}
	\centering
	\subfigure{
		\includegraphics[width=.34\textwidth]{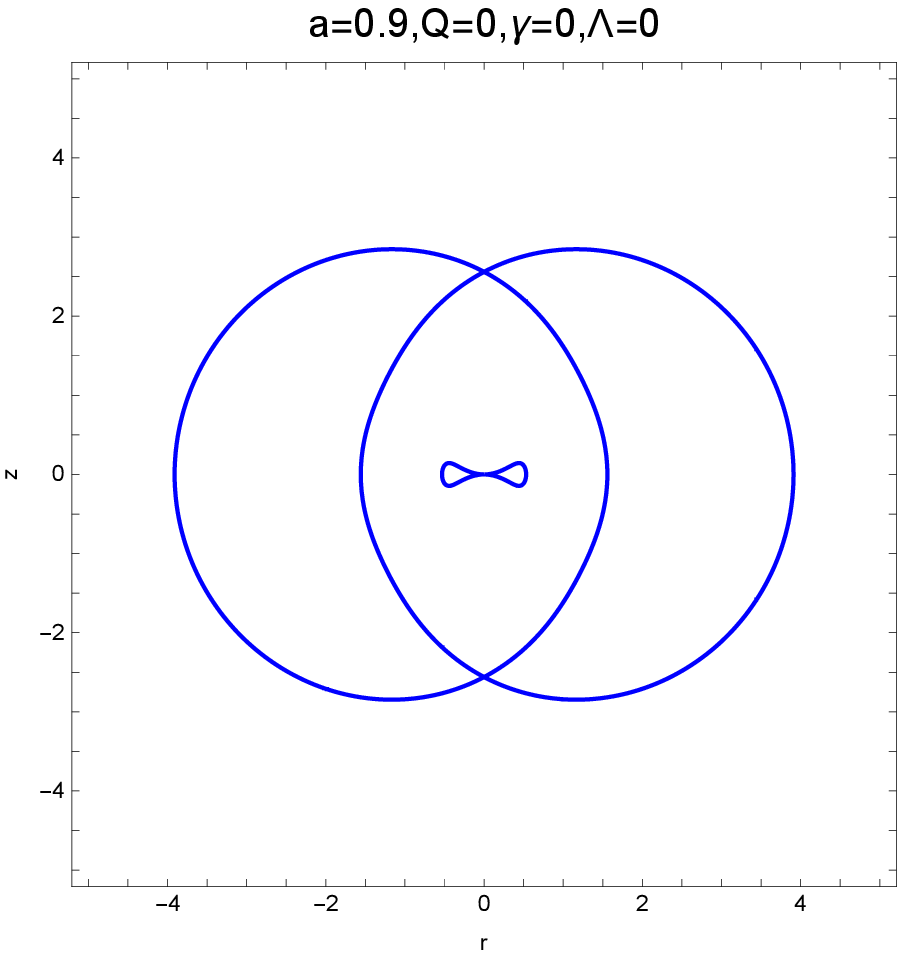}
		\includegraphics[width=.34\textwidth]{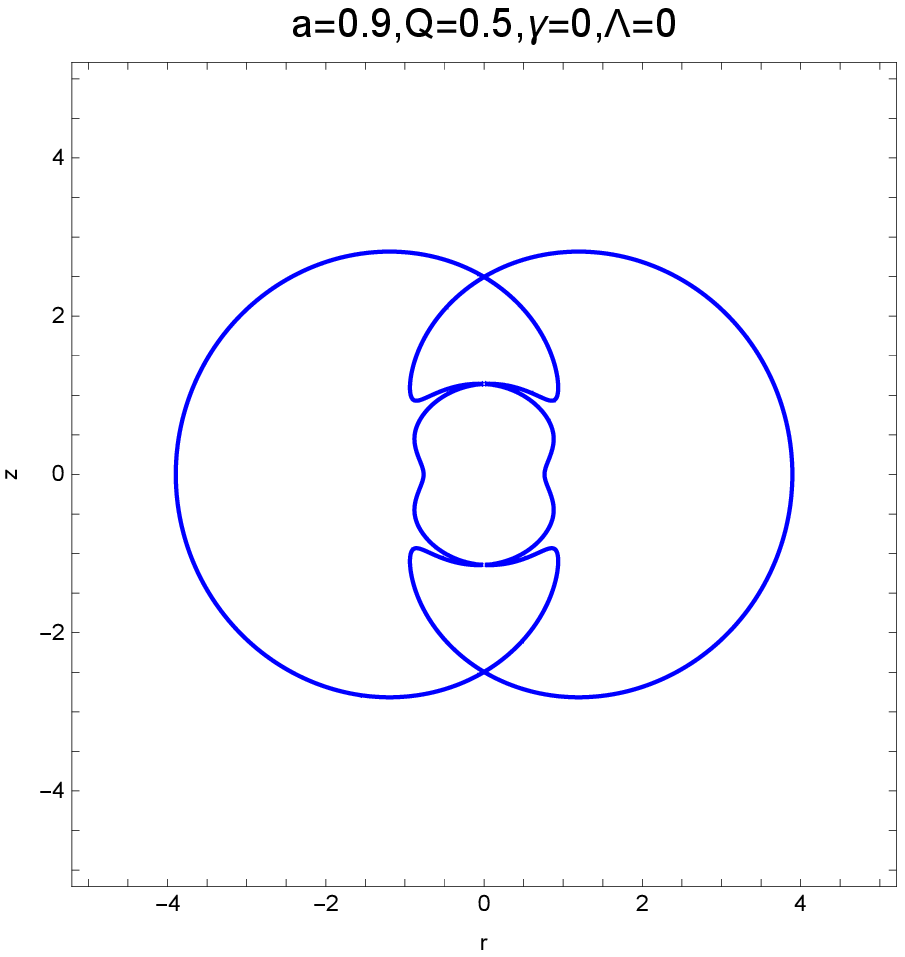}
		\includegraphics[width=.34\textwidth]{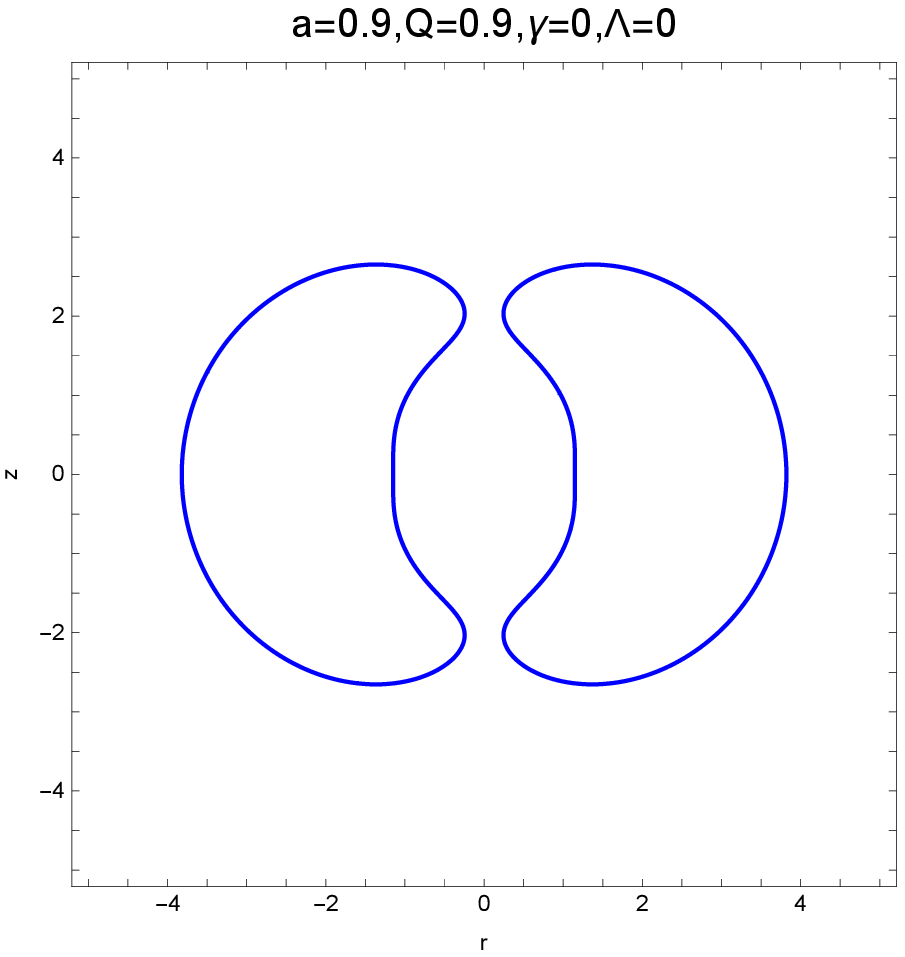}
	}
	\caption{Shape of photon regions with $\Lambda=0$ and $\gamma=0$ for a set of values of parameters $Q$ and $a$.}
	\label{photon1}
\end{figure}
By using the separation of variables method and introducing the Carter constant $K$, we can arrive at
 \begin{gather}
 \frac{\mathrm{d}S_{r}(r)}{\mathrm{d}r}=\frac{\sqrt{R(r)}}{\Delta_{r}},\label{eq:R}\\
\frac{\mathrm{d}S_{\theta}(\theta)}{\mathrm{d}\theta}=\frac{\sqrt{\Theta (\theta)}}{\Delta_{\theta}}\label{eq:O},
 \end{gather}  
  where   
 \begin{gather}
  	R(r)=(aL-(a^{2}+r^{2})E)^{2}\Xi^{2}-m^{2}r^{2}\Delta_{r}-K\Delta_{r},\label{eq:EM1}\\
  	\Theta(\theta)=K\Delta_{\theta}-a^{2}m^{2}\cos^{2}(\theta)-\Xi^{2}(L\csc(\theta)-aE\sin(\theta))^{2}.\label{eq:EM2}
  \end{gather}
 With these forms, we can rewrite the Hamilton-Jocobi function as
\begin{equation}\label{eq:hj2}
\begin{split}  
S=\frac{1}{2} m^{2}\tau -Et + L\phi + \int^{r}\frac{\sqrt{R(r)}}{\Delta_{r}}\mathrm{d}r+\int^{\theta}\frac{\sqrt{\Theta (\theta)}}{\Delta_{\theta}}\mathrm{d}\theta.
 \end{split}
  \end{equation}
Following \cite{carter}, we write geodesic motion in the form
\begin{gather}
 \Sigma\frac{\mathrm{d}t}{\mathrm{d}\tau}=\frac{\Xi^{2}((r^{2}+a^{2})E-aL)(r^{2}+a^{2})}{\Delta_{r}}-\frac{a\Xi^{2}(aE\sin^{2}(\theta)-L)}{\Delta_{\theta}},\label{eq:t}\\
\Sigma\frac{\mathrm{d}r}{\mathrm{d}\tau}=\sqrt{R(r)},\label{eq:r}\\
\Sigma\frac{\mathrm{d}\theta}{\mathrm{d}\tau}=\sqrt{\Theta (\theta)},\label{eq:o}\\
\Sigma\frac{\mathrm{d}\phi}{\mathrm{d}\tau}=\frac{a\Xi^{2}((r^{2}+a^{2})E-aL)}{\Delta_{r}}-\frac{\Xi^{2}(aE\sin^{2}(\theta)-L)}{\Delta_{\theta}\sin^{2}(\theta)}.\label{eq:f}
 \end{gather}  
 Here, we're interested in lightlike geodesics that stay on a sphere with a constant radiu. For lightlike geodesics, $m$ of test particle is zero. The region filled by these geodesics can be called as the photon region.

\begin{figure}[htp]
	\centering
	\includegraphics[width=.24\textwidth]{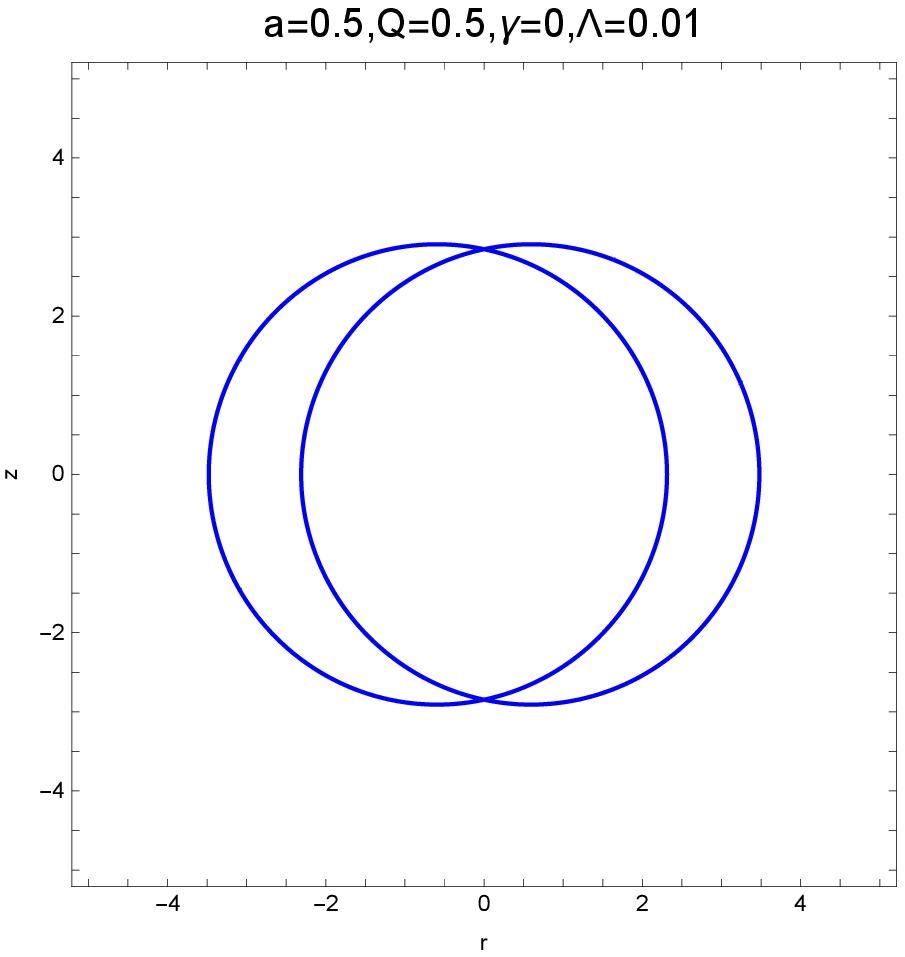}
	\includegraphics[width=.24\textwidth]{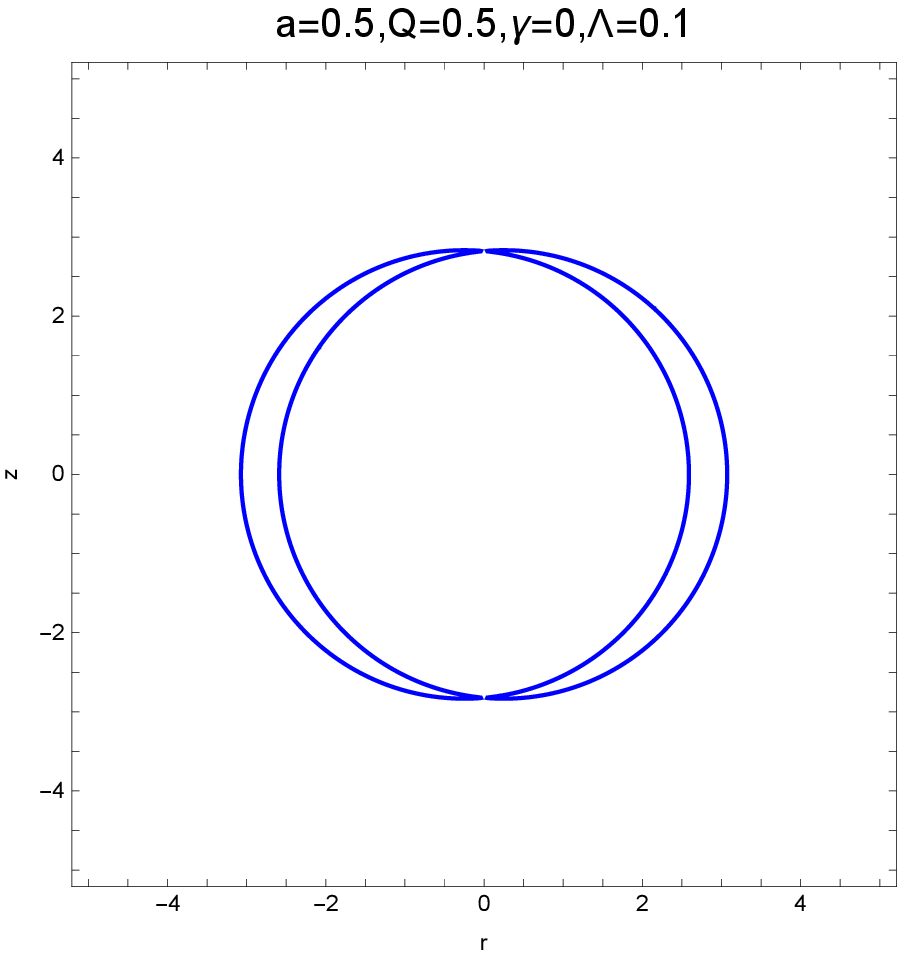}
	\includegraphics[width=.24\textwidth]{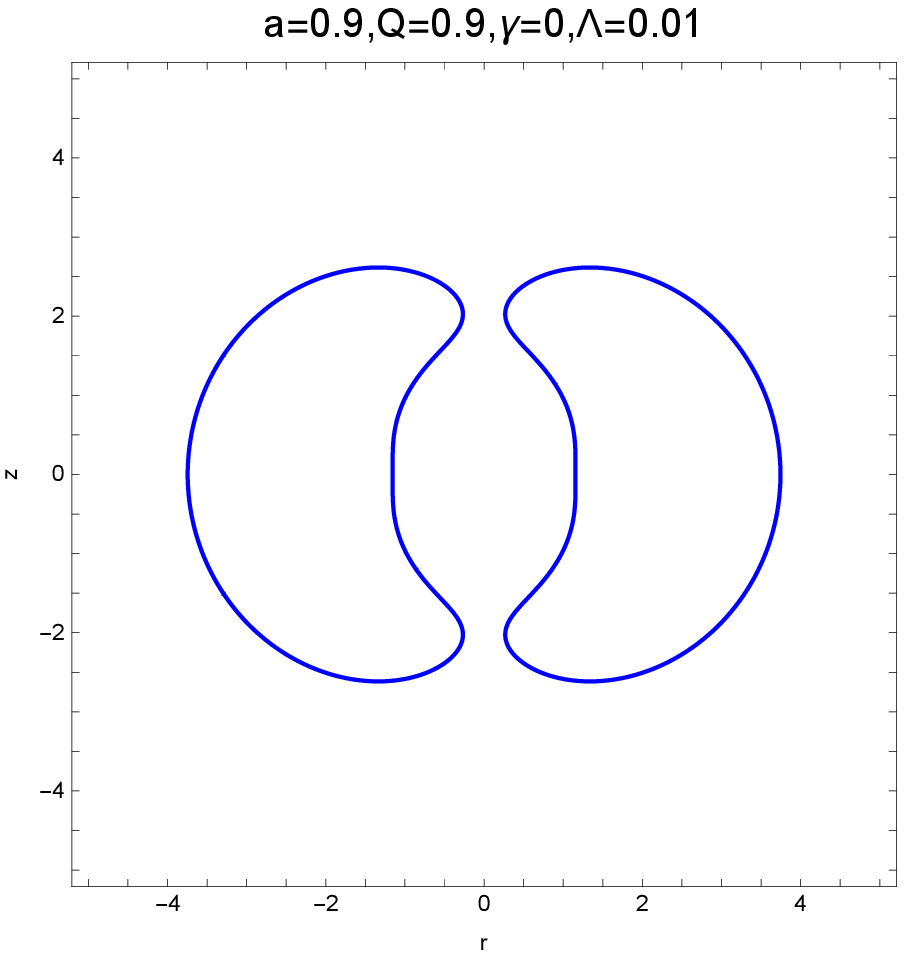}
	\includegraphics[width=.24\textwidth]{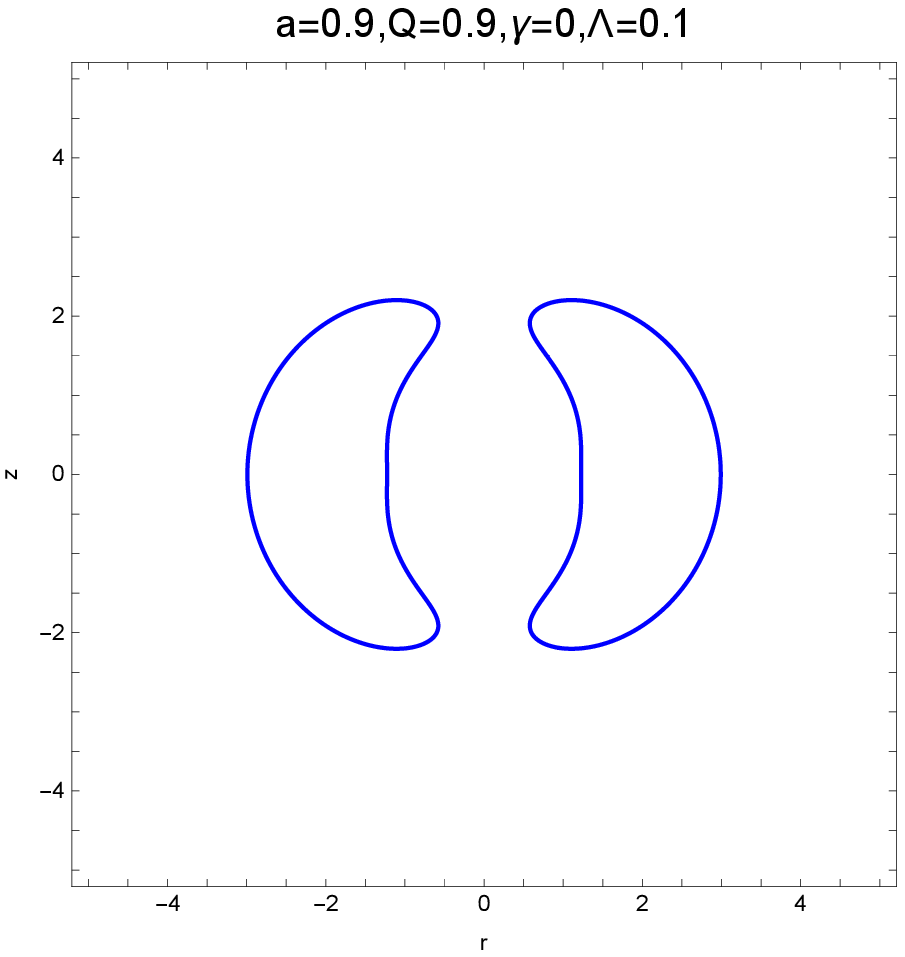}
	\caption{Shape of photon regions with fixed $a$ and $Q$ for a set of values of parameters $\Lambda$}
	\label{photon2}
\end{figure}

The conditions $\frac{\mathrm{d}r}{\mathrm{d}\tau}=0$ and  $\frac{\mathrm{d}^{2}r}{\mathrm{d}\tau^{2}}=0$ must be satisfied simultaneously to get spherical orbits, which means that 
\begin{equation}
R(r)=0 , \quad
 \frac{\mathrm{d}R(r)}{\mathrm{d}r}=0. 
\end{equation}
For simplicity, we introduce two reduced parameters $\xi=\frac{L}{E},\eta=\frac{K}{E^{2}}$. By solving above equations, we can obtain $\xi$ and $\eta$ for limiting spherical lightlike geodesic, by
\begin{gather}
 a\xi=r_{p}^{2}+a^{2}-4\frac{r \Delta_{r}}{\Delta{'}_{r}}\label{eq:R1},\\
\eta =\frac{16r^{2}\Xi^{2}\Delta_{r}}{(\Delta'_{r})^{2}}\label{eq:O1}.
 \end{gather} 
 Inserting (\ref{eq:R1}) and (\ref{eq:O1}) into $\Theta\geq0$, we find the condition which describes the photon region. We will discuss th photon regions in $(r,\theta)$ plane
\begin{equation}\label{eq:ph}
16r^{2}a^{2}\Delta_{r}\Delta_{\theta}\sin^{2}(\theta)-(4r\Delta_{r}-\Sigma\Delta'_{r})^{2}\geq 0.
\end{equation}
And we have shown the shape of photon region with different values of $a$, $Q$, $\gamma$ and $\Lambda$ in Fig.~\ref{photon1} and Fig.~\ref{photon2}. From the picture, we can see some interesting phenomenon. For example, when $a=0$, Eq. (\ref{eq:ph}) reduces to 
\begin{equation}\label{eq:ph0}
(4r\Delta_{r}-\Sigma\Delta'_{r})^{2}= 0.
\end{equation}
And it means that photon regions reduce to photon spheres. In Schwarzschild space-time, i.e , $a$, $Q$, $\gamma$ and $\Lambda$ become zero simultaneously, the photon sphere stay at $r=3$. When $a=0$ with a non-zero $Q$, there will be two photon spheres. And the radiu of inner photon sphere increase as $Q$ increases. With increasing $a$, the photon sphere will expand to two crescent-shaped axisymmetric region in $(r,\theta)$ plane and a interior photon region will rise. And non-zero $Q$ will diminish the area enclosed by the two crescent-shaped photon region. Keeping increasing $Q$, the exterior photon regions will touch the interior photon region, get deformation with the disappearance of interior regions. For a large $Q$, photon regions will separate into two untouched regions. From Fig.~\ref{photon2}, we can see that photon regions will shrink with a increasing $\Lambda$. For quintessential parameter $\gamma$, it gets more interesting, as there can be two different situation. 
\par
   \begin{figure}[htp]
 	\centering
 	\includegraphics[width=.24\textwidth]{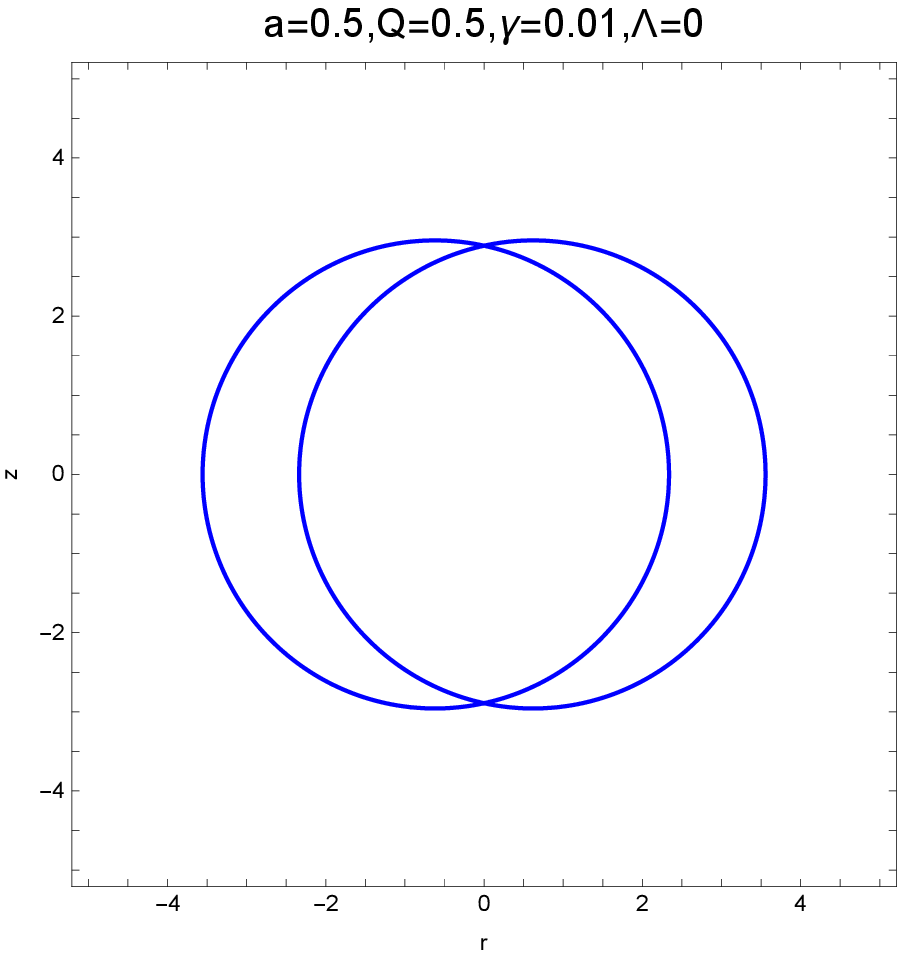}
 	\includegraphics[width=.24\textwidth]{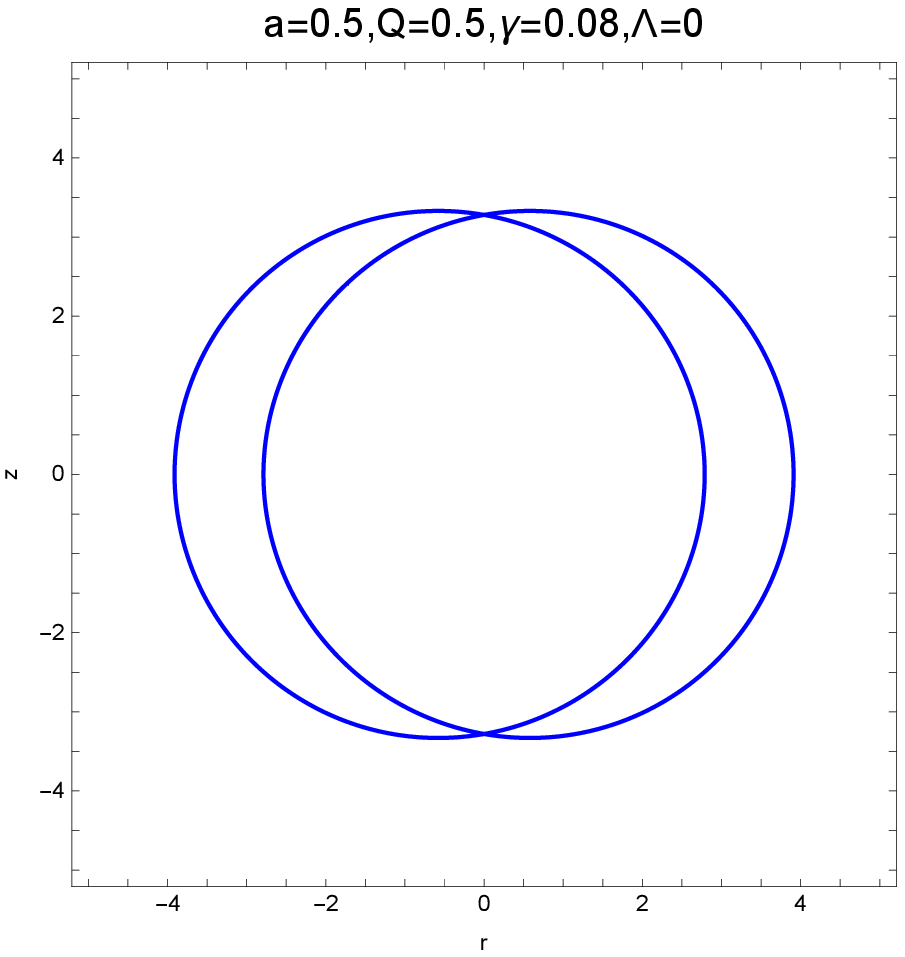}
 	\includegraphics[width=.24\textwidth]{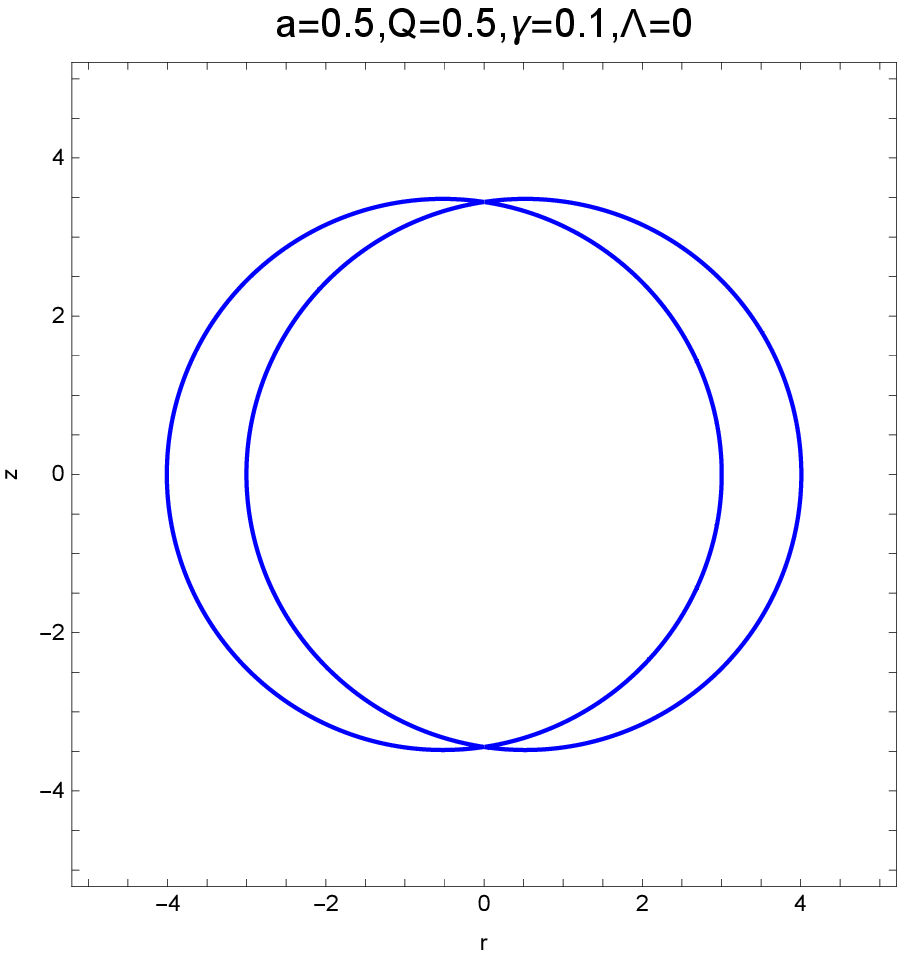}
 	\includegraphics[width=.24\textwidth]{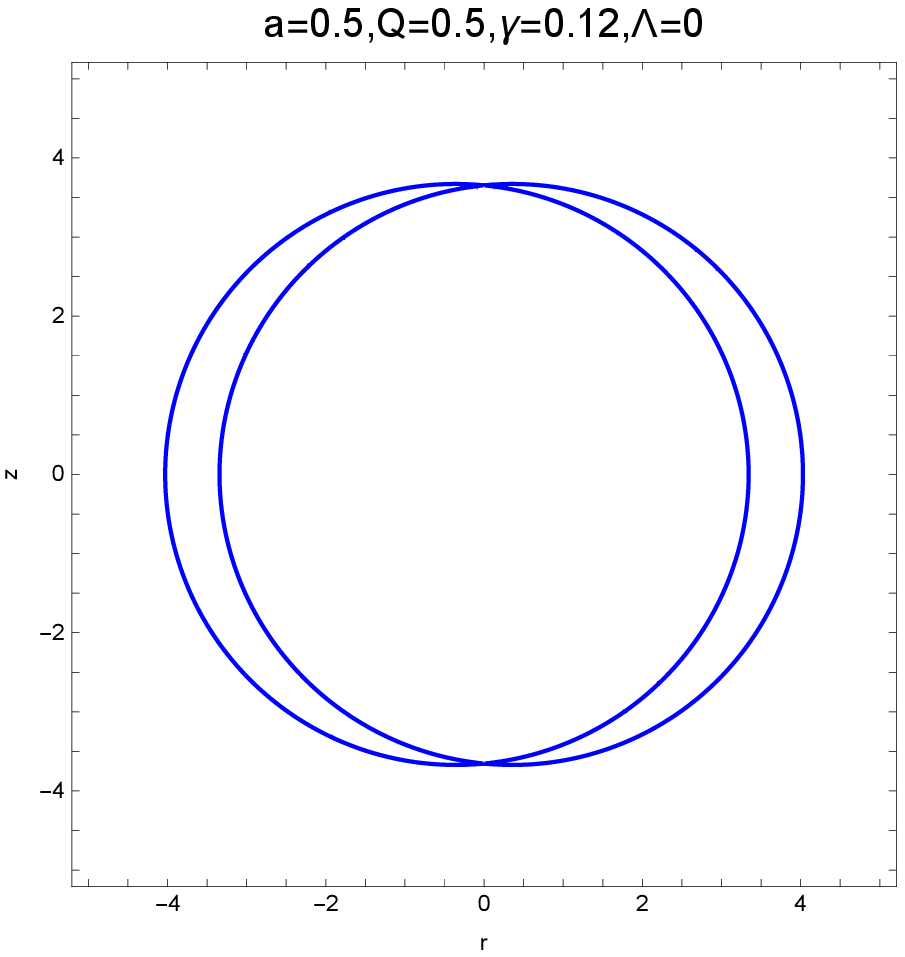}
 	\includegraphics[width=.24\textwidth]{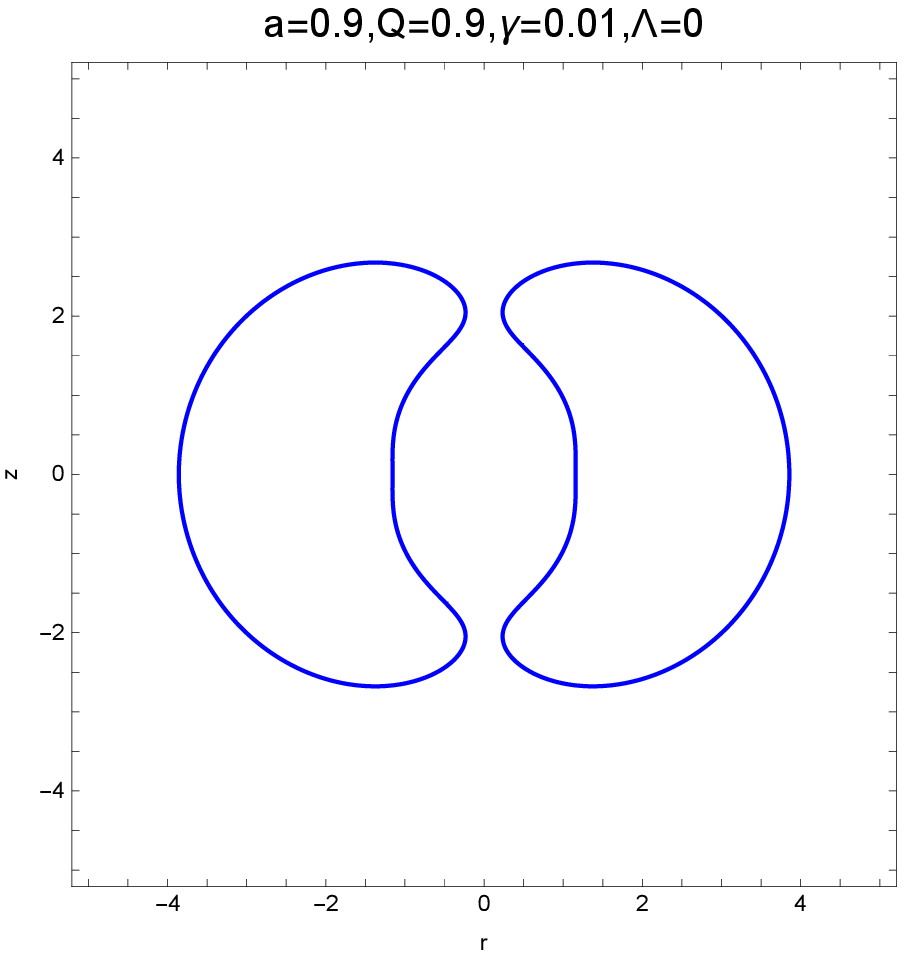}
 	\includegraphics[width=.24\textwidth]{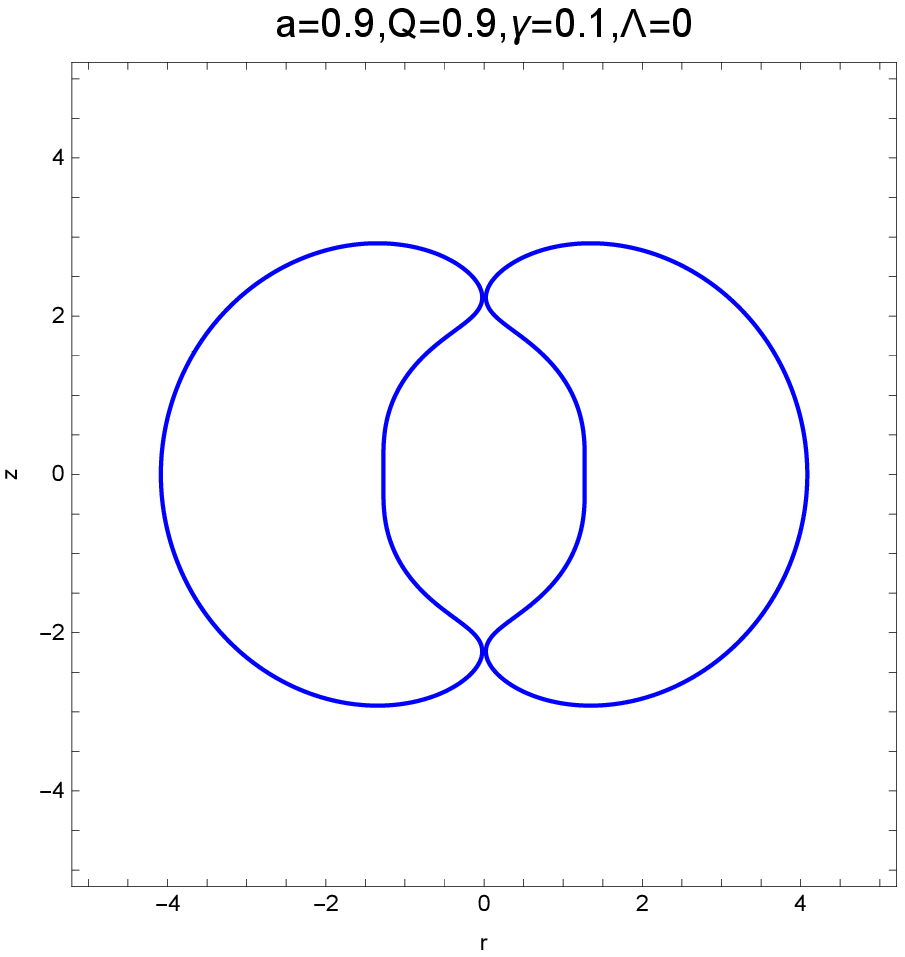}
 	\includegraphics[width=.24\textwidth]{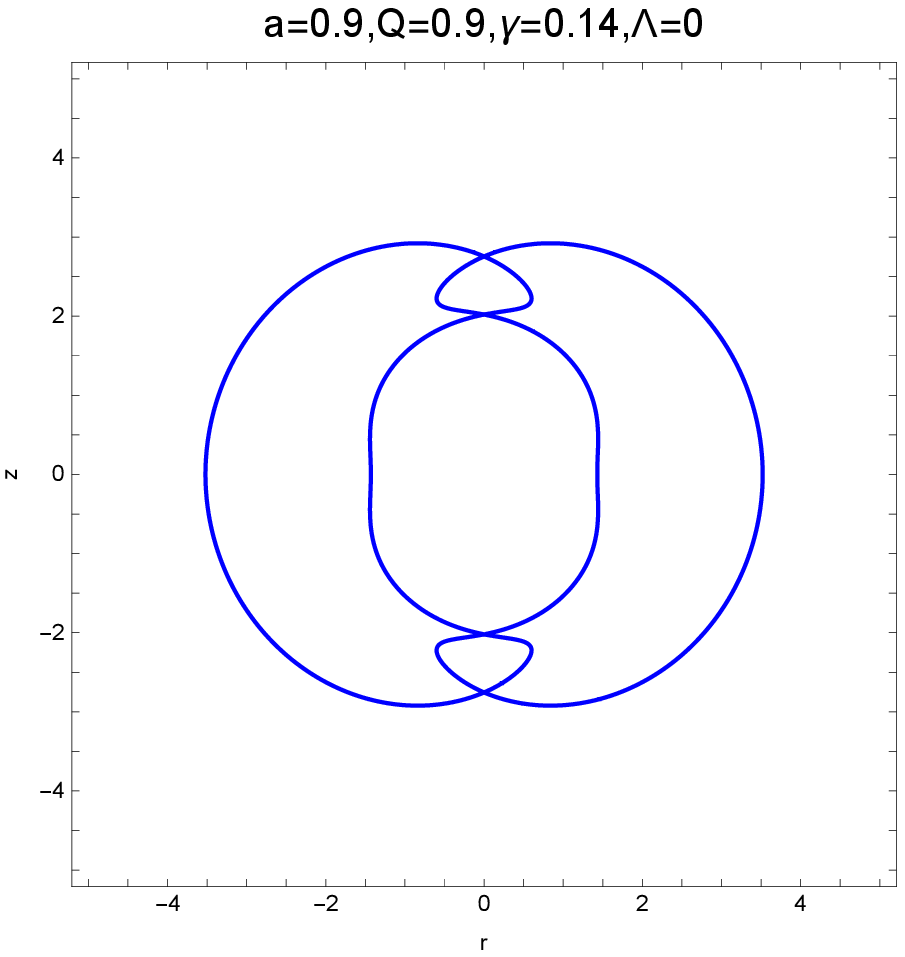}
 	\includegraphics[width=.24\textwidth]{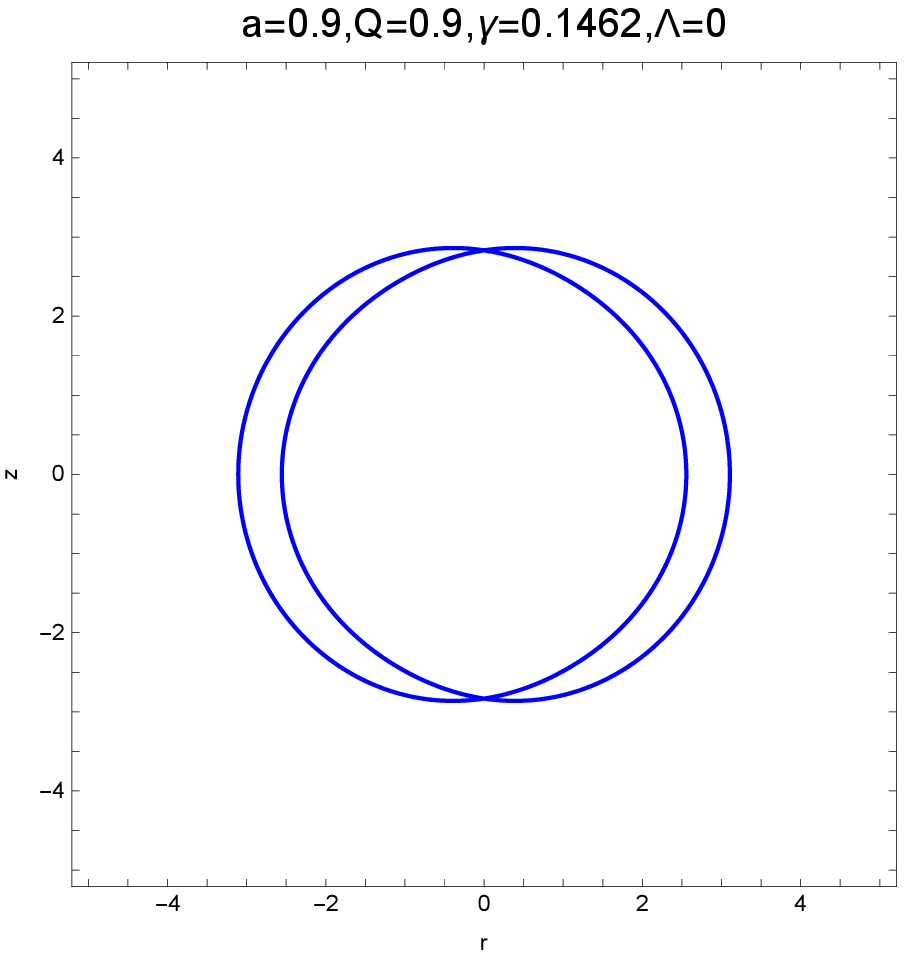}
 	
 	\caption{Shape of photon regions with fixed $a$ and $Q$ for a set of values of parameters $\gamma$}
 	\label{photon3}
 \end{figure}
 $Type{I}$. For a small $a$ and $Q$, two exterior photon regions maintain crescent shape and keep touching each other. Increasing value of $\gamma$ will stretch the photon regions to both sides and diminish the regions. See the upper row in Fig.~\ref{photon3}.
\par
$Type{II}$. For a large $a$ and $Q$, the two regions get deformation like orange segments and separated from each other. With a increasing $\gamma$, the two separated regions will grow until touching each other and take a deformation. If we keep pushing up the value of $\gamma$, the two regions will go back to crescent shape and shrink, as shown in the lower row of Fig.~\ref{photon3}.

 \section{shadow}\label{6}
 A general method for computing shadows of black hole is that locate observers at a place near infinitely away from the black hole. And the shadow can be determined from the viewpoint of the observer. For a asymptotically flat spacetime, this method can be effective, but considering a non-zero cosmological constant, the position of the observer should be fixed. So for nonasymptotically flat spacetime, there is a difficulty that the method that locate the source and observer at infinity is no longer feasible. So in this paper, by following \cite{grenzebach2014photon,haroon2019shadow} we use a different method to compute the shadows.
 \par
  At first, we consider the observer at a fixed location $(r_{0},\theta_{0})$ where $\Delta_{r}>0$ in the Boyer-Lindquist coordinates. And then we consider light rays sent from the location  $(r_{0},\theta_{0})$ to the past. So we can characterized such lightlike geodesics into two different categories. For the first type, they get so close to the outer horizon that they absorbed by the gravity, so the light rays cannot come from our source. And what the observer can see is darkness in these directions. And for the other one, they avoid being absorbed and in these direction it is bright for the observer. And the boundary of the shadow can be get from borderline case.
  \par
  Now, we chose an orthonormal tetrad 
\begin{equation}\label{eq:basis}
\begin{split}
e_{0}=&\frac{\Xi}{\sqrt{\Delta_{r}\Sigma}}\left(\left(r^{2}+a^{2}\right)\partial_{t} + a\partial_{\phi}\right)\big|_{(r_{0},\theta_{0})},\\
e_{1}=&\sqrt{\frac{\Delta_{\theta}}{\Sigma}}\partial_{\theta}\big|_{(r_{0},\theta_{0})},\\
e_{2}=&-\frac{\Xi}{\sqrt{\Delta_{\theta}\Sigma\sin^{2}{\theta}}}\left(a\sin^{2}{\theta}\partial_{t}+\partial_{\phi}\right)\big|_{(r_{0},\theta_{0})},\\
e_{3}=&-\sqrt{\frac{\Delta_{r}}{\Sigma}}\partial_{r}\big|_{(r_{0},\theta_{0})},
\end{split}
\end{equation} 
where the $e_{0}$ represents the four-velocity of the observer, $e_{0}\pm e_{3}$ are
tangent to the principal null congruence and $e_{3}$ is along the direction towards the black hole's center. And the coordinates of light rays $\lambda(\tau)$ can be written in $(t(\tau),r(\tau),\theta(\tau),\phi(\tau))$, the tangent vector can be given as
\begin{equation}\label{tan1}
\dot{\lambda}=\dot{t}\partial_{t}+\dot{r}\partial_{r}+\dot{\theta}\partial_{\theta}+\dot{\phi}\partial_{\phi}.
\end{equation}
And this tangent vector at observer event can also be described  as
\begin{equation}\label{tan2}
\dot{\lambda} = \alpha(-e_{0}+\sin\Upsilon\cos\Phi e_{1}+\sin\Upsilon \sin\Phi e_{2}+\cos\Upsilon e_{3}).
\end{equation}
\par
From Eq.~(\ref{eq:t})-(\ref{eq:f}) and Eq.~(\ref{eq:basis}), the scalar factor $\alpha$ can be obtain by 
\begin{equation}
\alpha=g(\dot{\lambda},e_{0})=\Xi\frac{aL-E(r^{2}+a^{2})}{\sqrt{\Delta_{r}\Sigma}}\big|_{(r_{0},\theta_{0})}.
\end{equation}
By comparing Eq.~(\ref{eq:t})-(\ref{eq:f}) and Eq.~(\ref{eq:basis}), the coordinates $\Upsilon$ and $\Phi$ can be defined in terms od $\xi$ and $\eta$. With the help of
\begin{equation}
\begin{split}
R=E^{2}(\Xi^{2}(r^{2}+a^{2}-a\xi)^{2}-\Delta_{r}\eta),\\
\Theta=E^{2}(\eta\Delta_{\theta}-\Xi^{2}(a\sin(\theta)-\xi\csc(\theta)^{2})),
\end{split}
\end{equation}
one can obtain 
\begin{equation}\label{curve}
\begin{split}
\cos\Upsilon=\frac{\Sigma\dot{r}}{\Xi E((r^{2}+a^{2})-a\xi)}\big|_{(r_{0},\theta_{0})},\\ 
\sin\Phi=\frac{\sqrt{\Delta_{r}}\sin\theta}{\sqrt{\Delta_{\theta}}\sin\Upsilon}[\frac{a-\xi\csc^{2}\theta}{a\xi-(r^{2}+a^{2})}]\big|_{(r_{0},\theta_{0})}.
\end{split}
\end{equation}
\par
 \begin{figure}[htp]
	\centering
	\includegraphics[width=.32\textwidth]{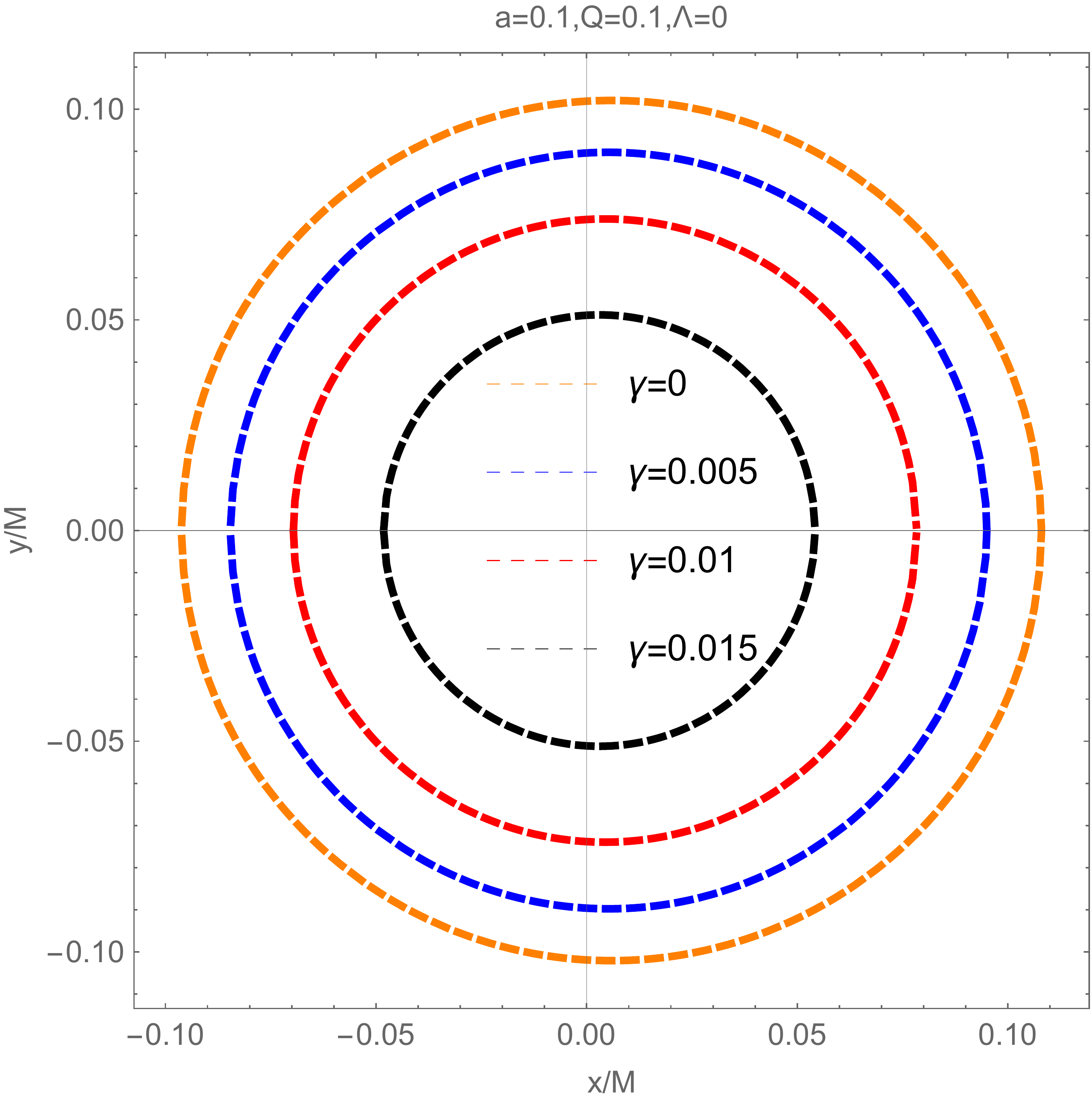}
	\includegraphics[width=.32\textwidth]{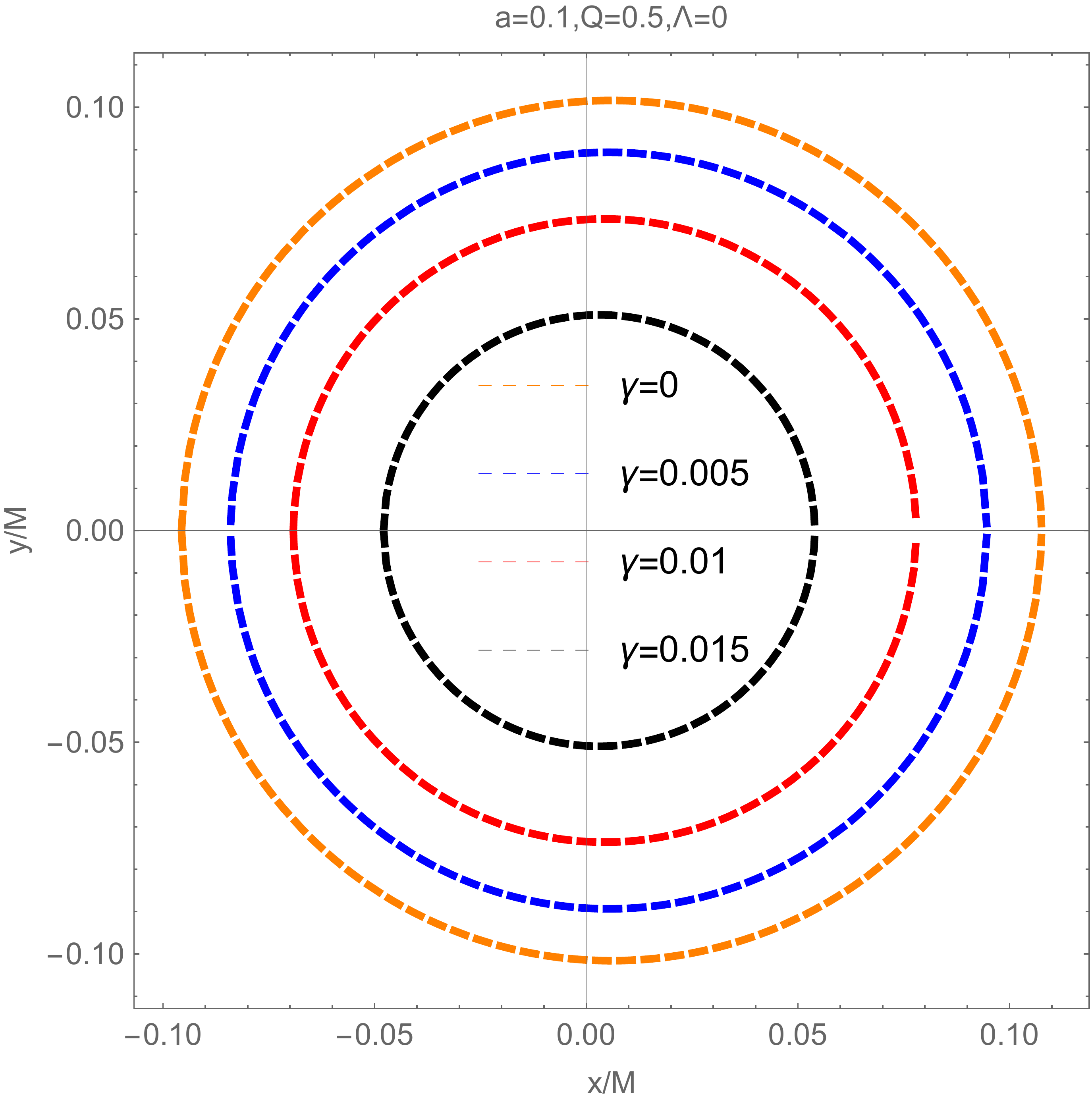}
	\includegraphics[width=.32\textwidth]{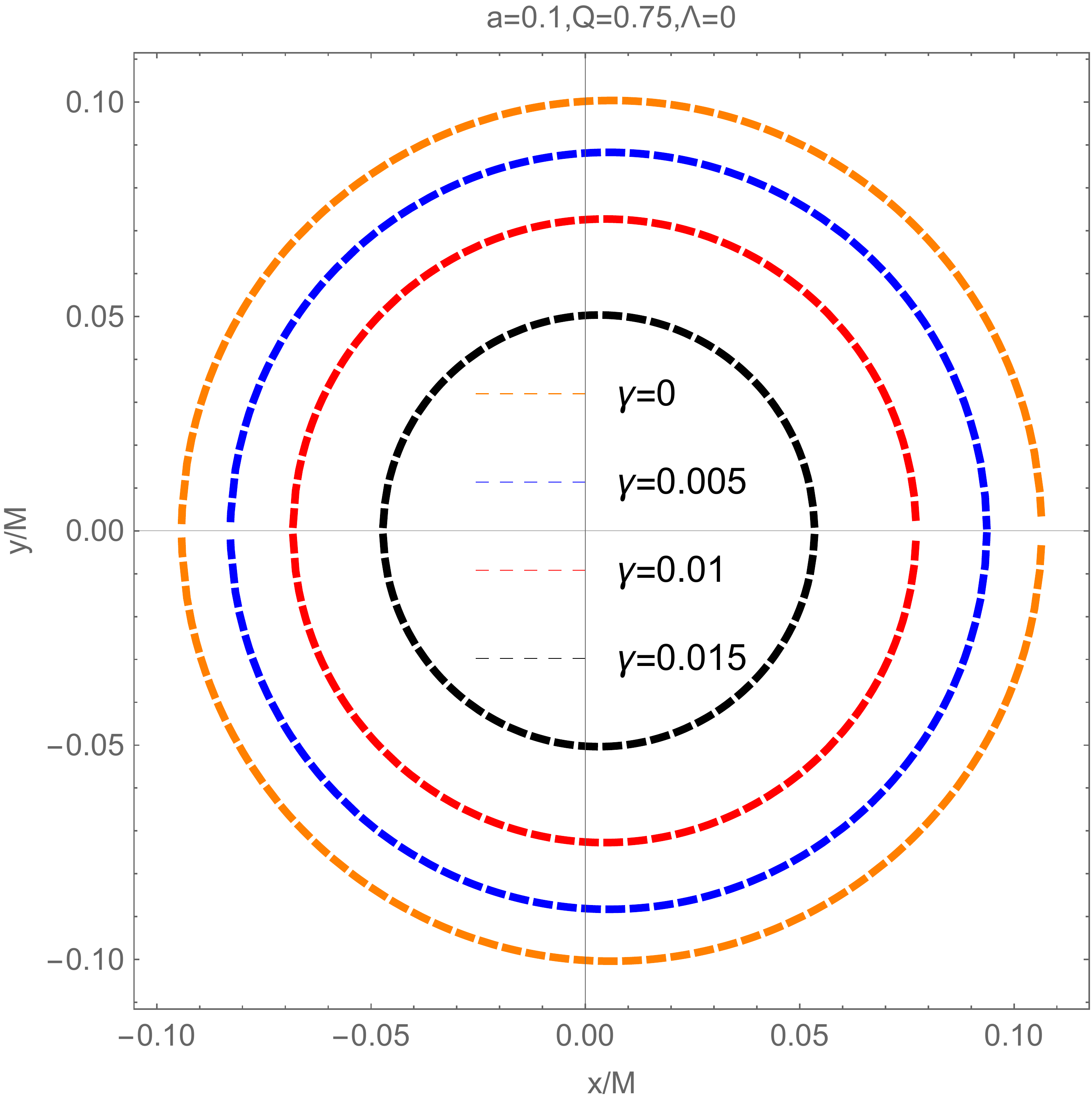}
	\includegraphics[width=.32\textwidth]{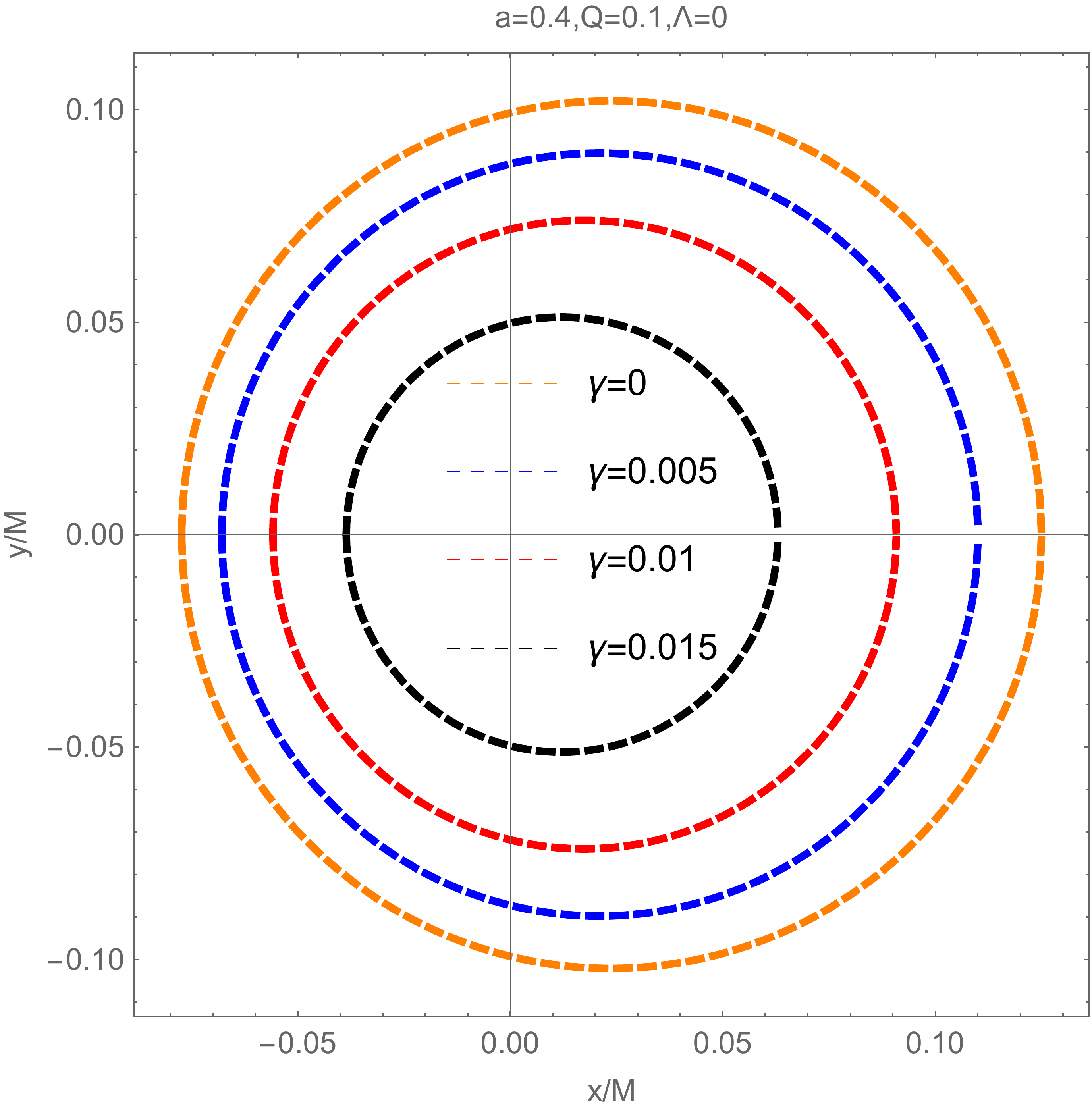}
	\includegraphics[width=.32\textwidth]{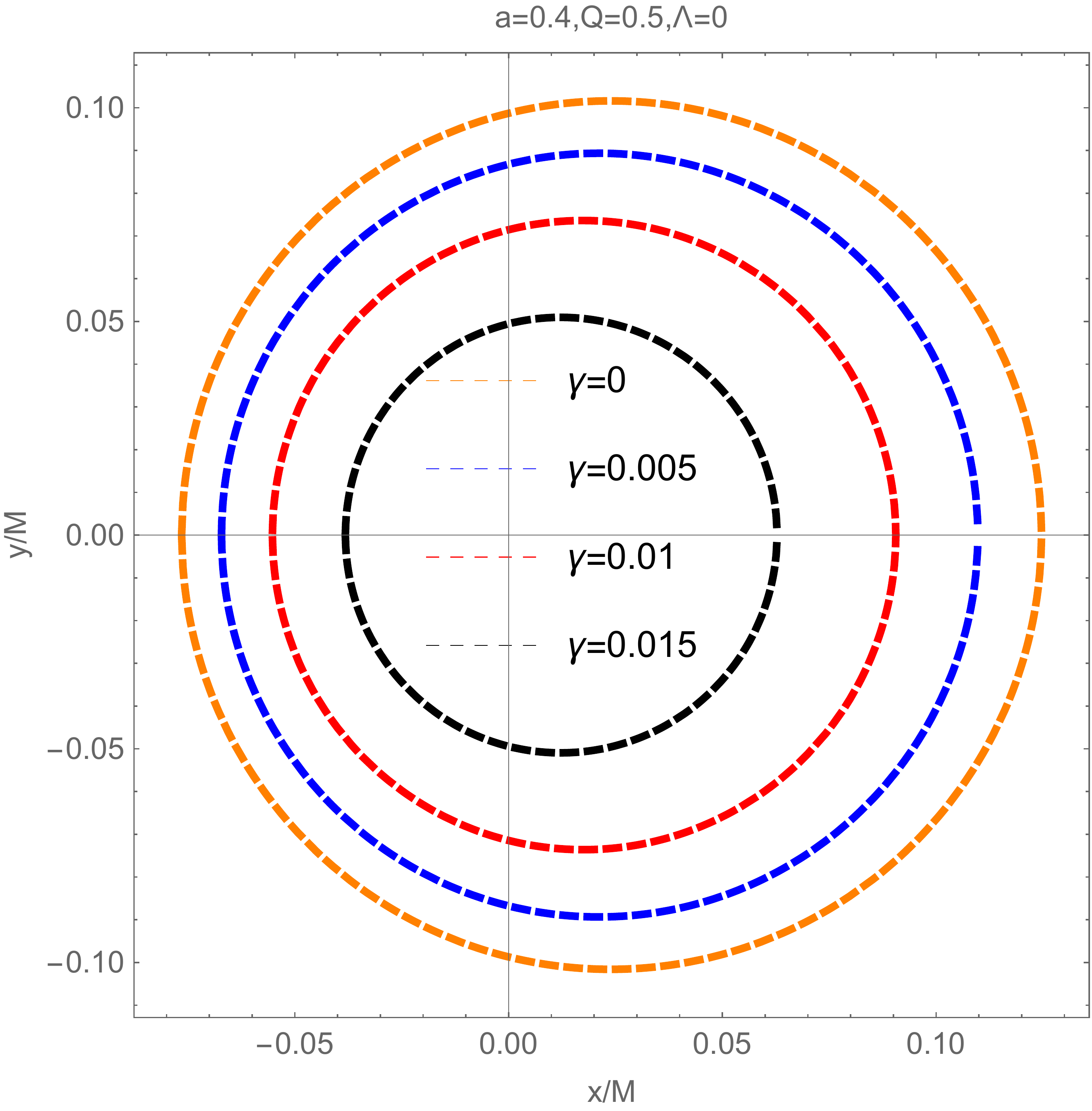}
	\includegraphics[width=.32\textwidth]{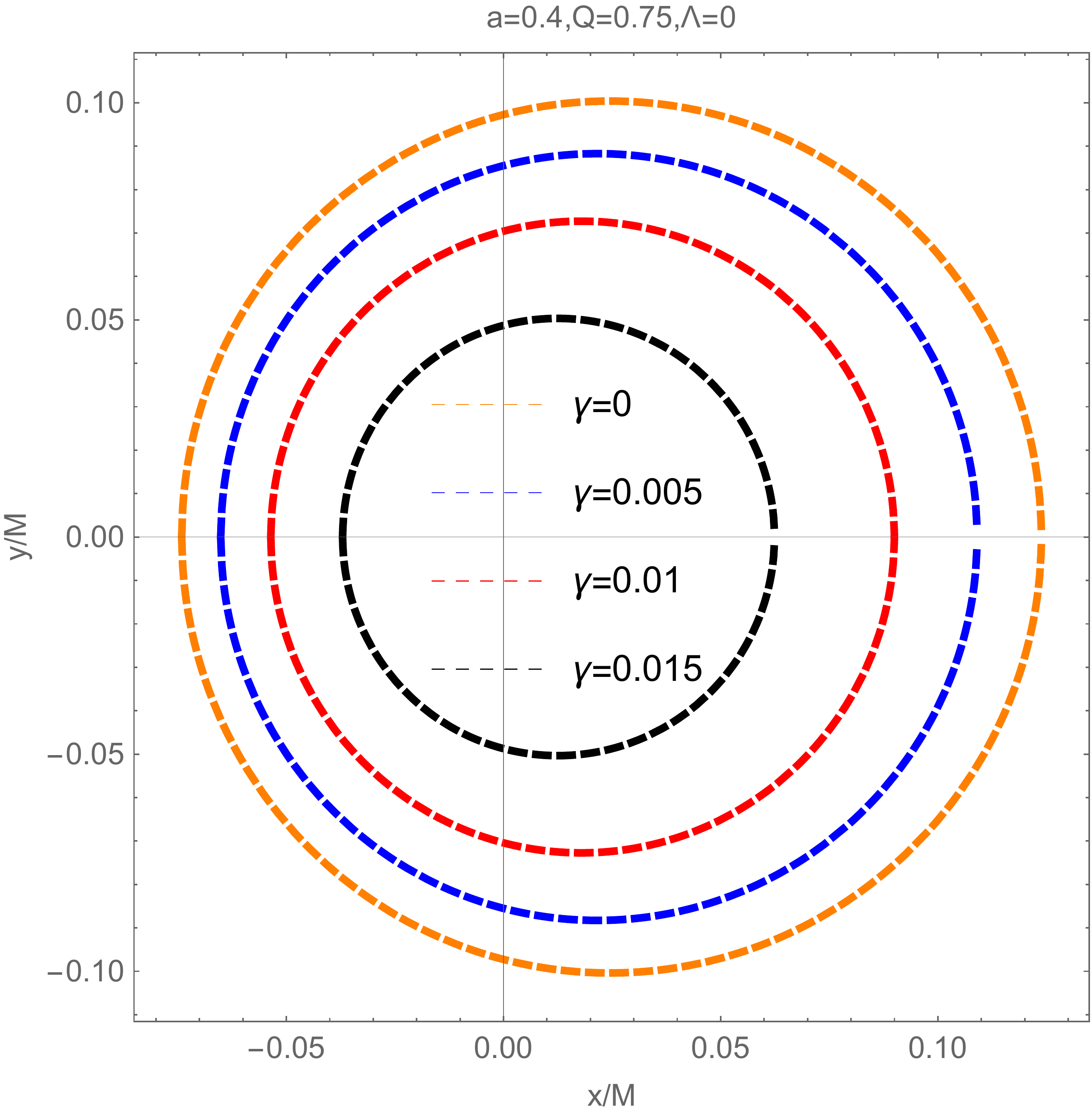}
	\includegraphics[width=.32\textwidth]{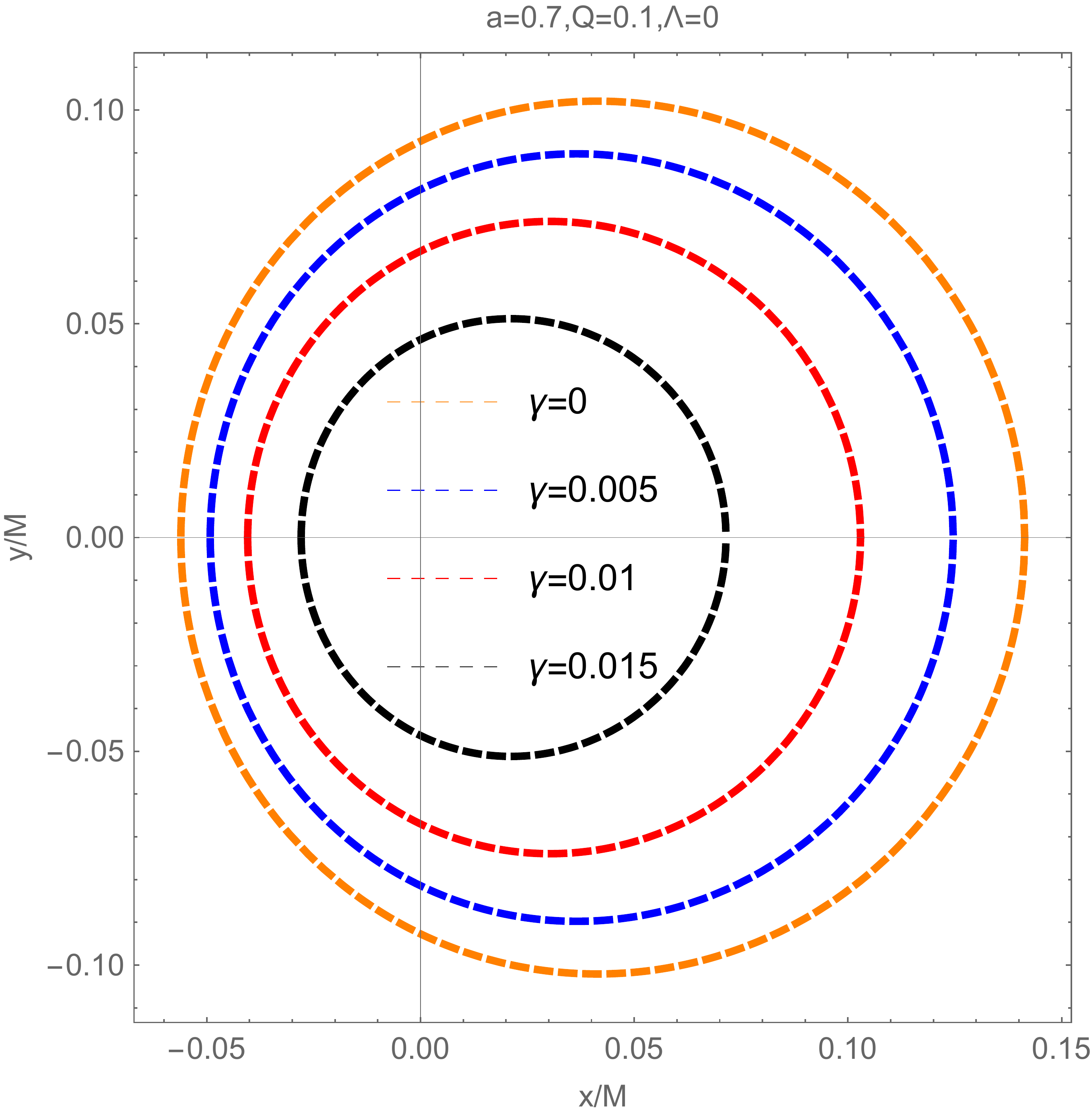}
	\includegraphics[width=.32\textwidth]{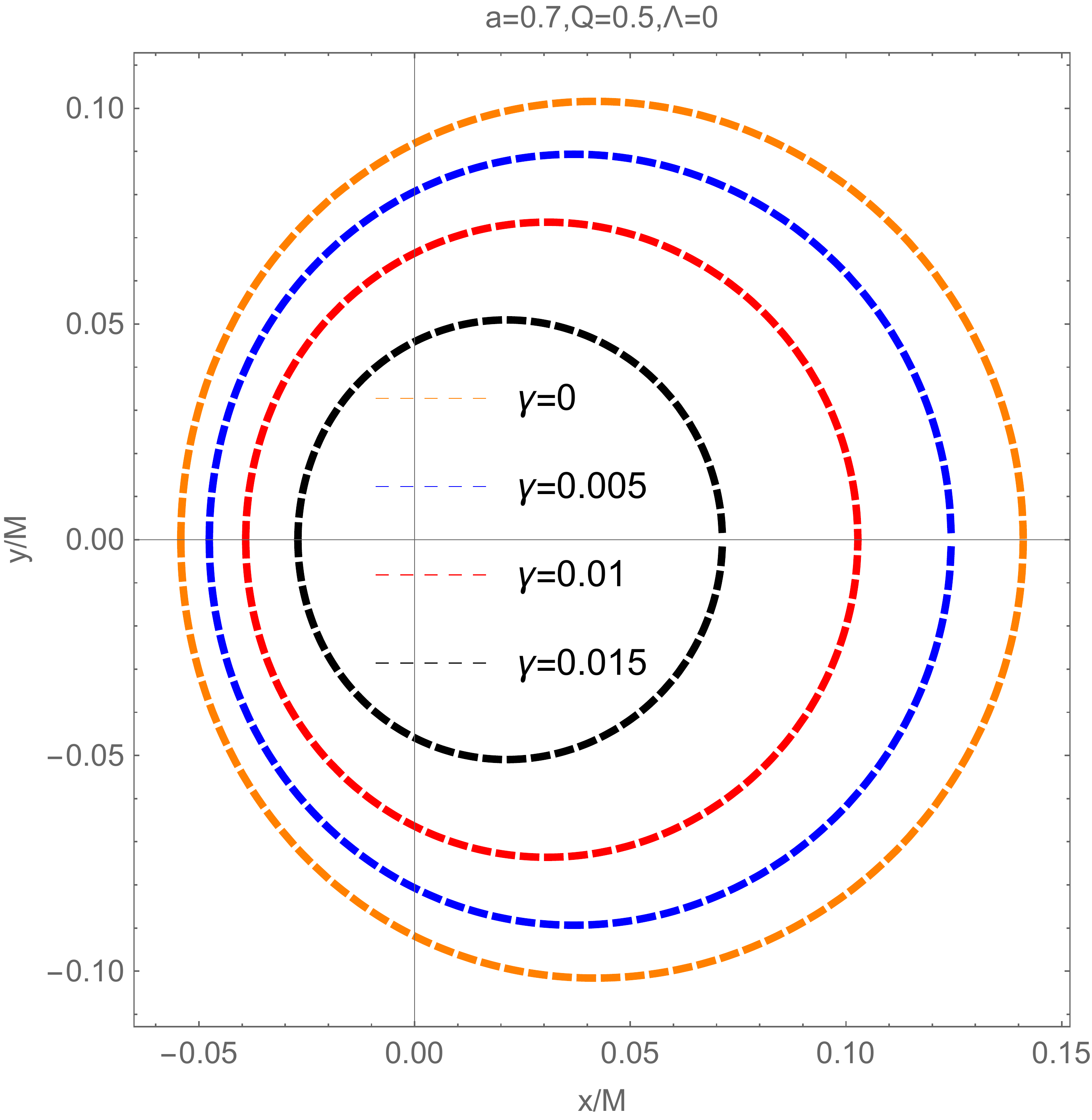}
	\includegraphics[width=.32\textwidth]{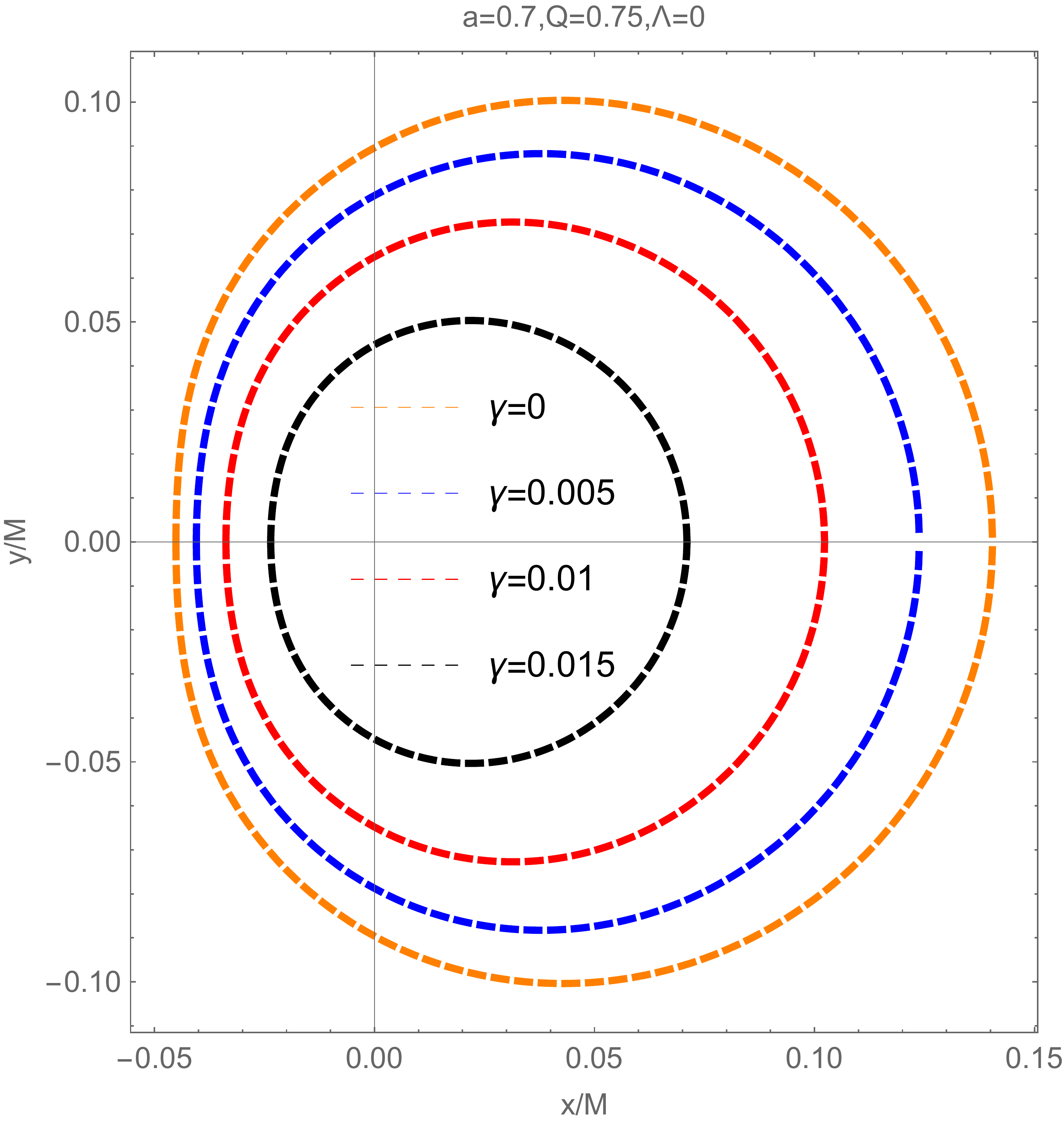}
	\caption{Shape of shadows with fixed $a$, $Q$, $\Lambda$ for a set of values of parameters $\gamma$.}
	\label{shadow1}
\end{figure}

 The shadow's boundary curve is determined by the light rays that asymptotically approach a spherical lightlike geodesic. So one has 
 \begin{gather}
 a\xi(r_{p})=r^{2}_{p}+a^{2}-4\frac{r_{p} \Delta_{r}(r_{p})}{\Delta{'}_{r}(r_{p})}\label{eq:R2},\\
 \eta(r_{p}) =\frac{16r^{2}_{p}\Xi^{2}\Delta_{r}(r_{p})}{(\Delta'_{r}(r_{p}))^{2}}\label{eq:O2},
 \end{gather} 
and $r_{p}$ represents the radiu coordinate of the limiting spherical lightlike geodesic. The boundary curve $(\Upsilon(r_{p}),\Phi(r_{p}))$ can therefore be given by inserting Eqs. (\ref{eq:R2}) and (\ref{eq:O2}) to Eq.(\ref{curve}).
\par
 To plot the shadows, we use a stereographic projection from the celestial sphere onto to a plane with the Cartesian coordinates. And for simplicity, we set $\theta_{0}=\frac{\pi}{2}$ and $r_{0}=50$ in our work.  
 \begin{gather}
 x=-2\tan(\frac{\Upsilon}{2})\sin\Phi,\\
 y=-2\tan(\frac{\Upsilon}{2})\cos\Phi.
 \end{gather}
 We show the shape of shadows in different values of $a$ and $Q$ with a set of  quintessence parameter $\gamma$ in Fig.~\ref{shadow1}.  

From Fig.~\ref{shadow1}, we can see that the spin of black hole will elongates the shadow and distorts the shadow with a increasing $a$. On the other hand, with a fixed $a$, the nonlinear magnetic charge $Q$ will decrease the size of shadow and can also distorts the shadow. And the intensity of quintessence field will diminish the size of shadows.
\par
 \begin{figure}[htp]
	\centering
	\includegraphics[width=.24\textwidth]{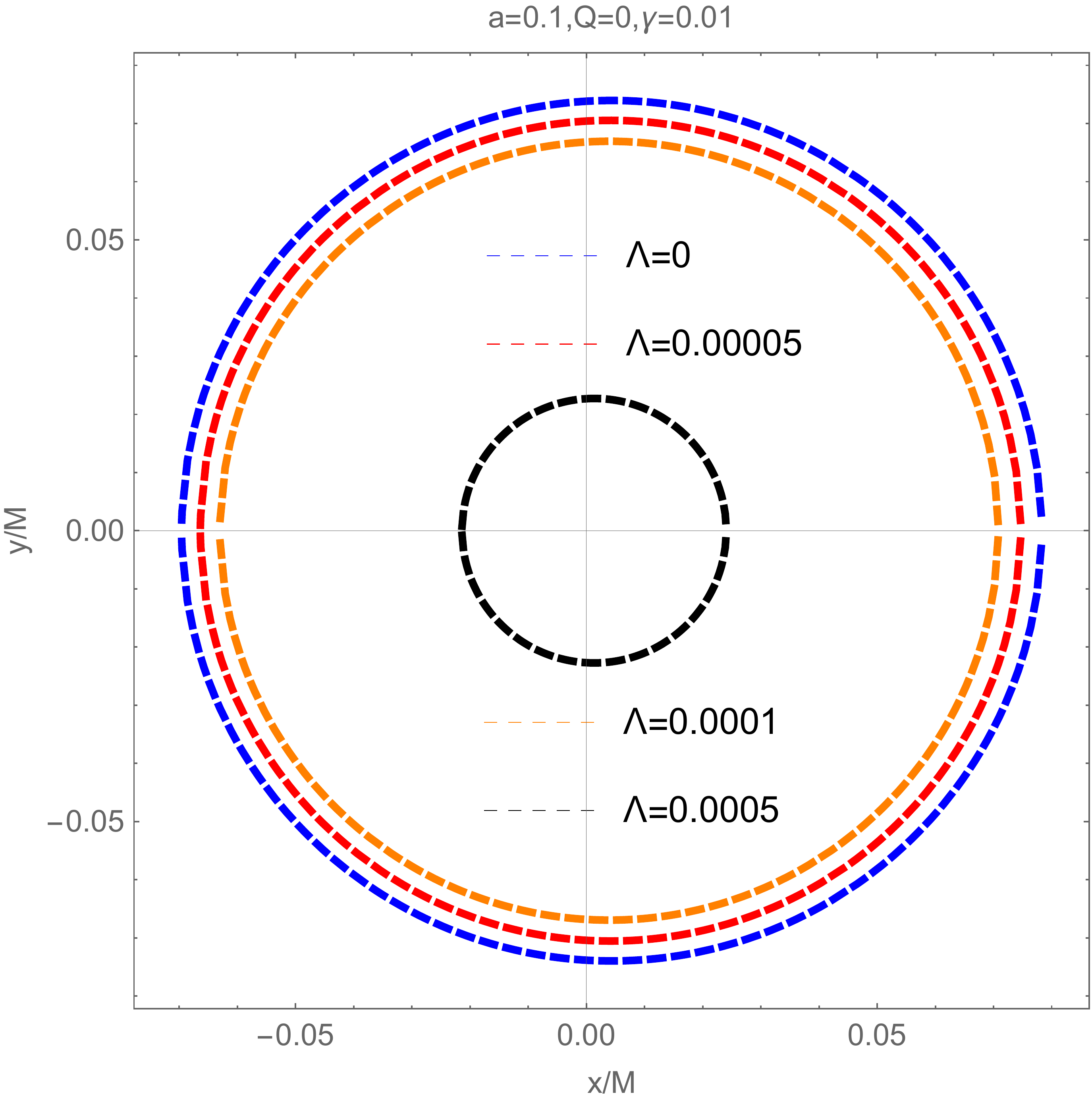}
	\includegraphics[width=.24\textwidth]{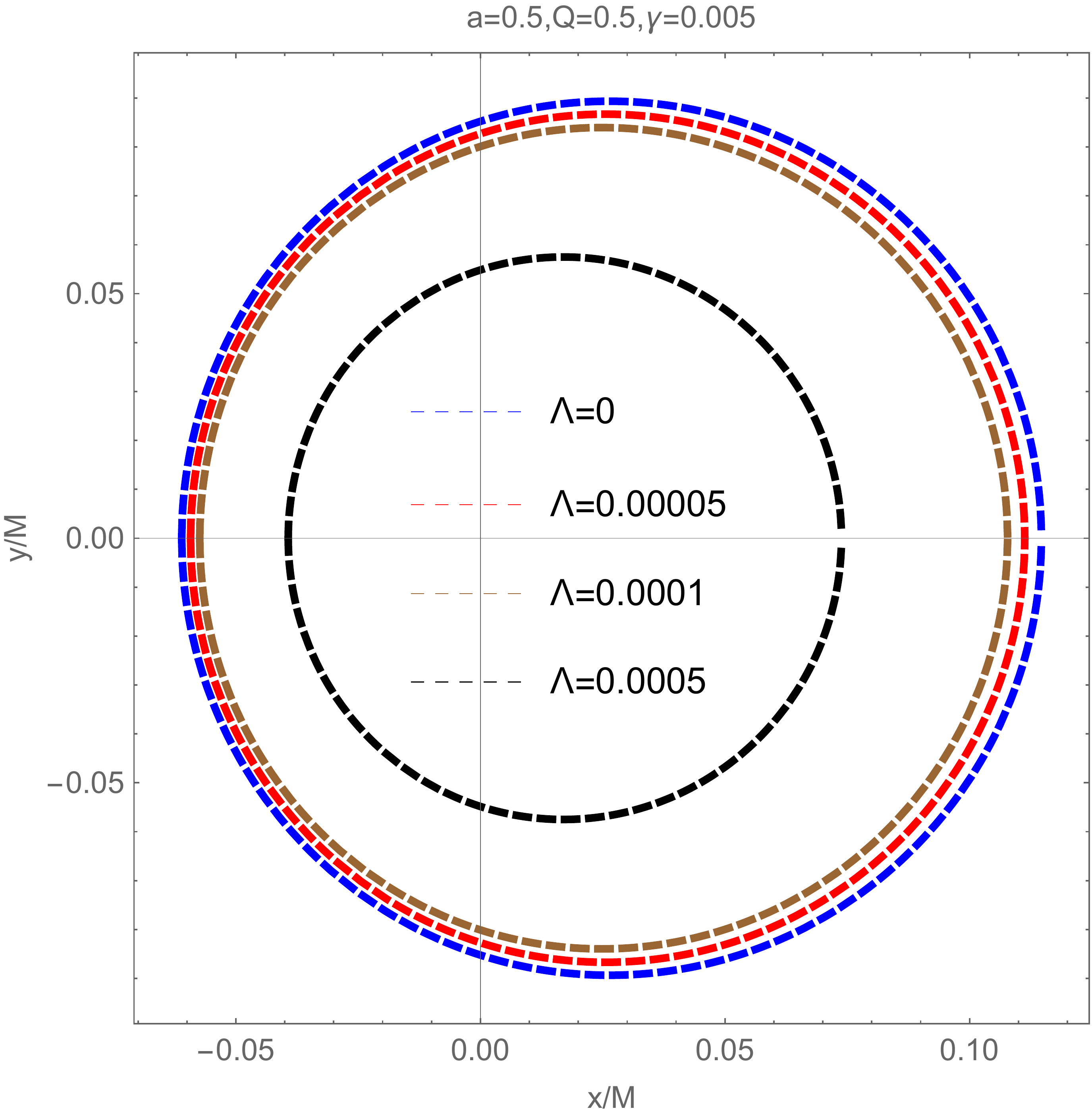}
	\includegraphics[width=.24\textwidth]{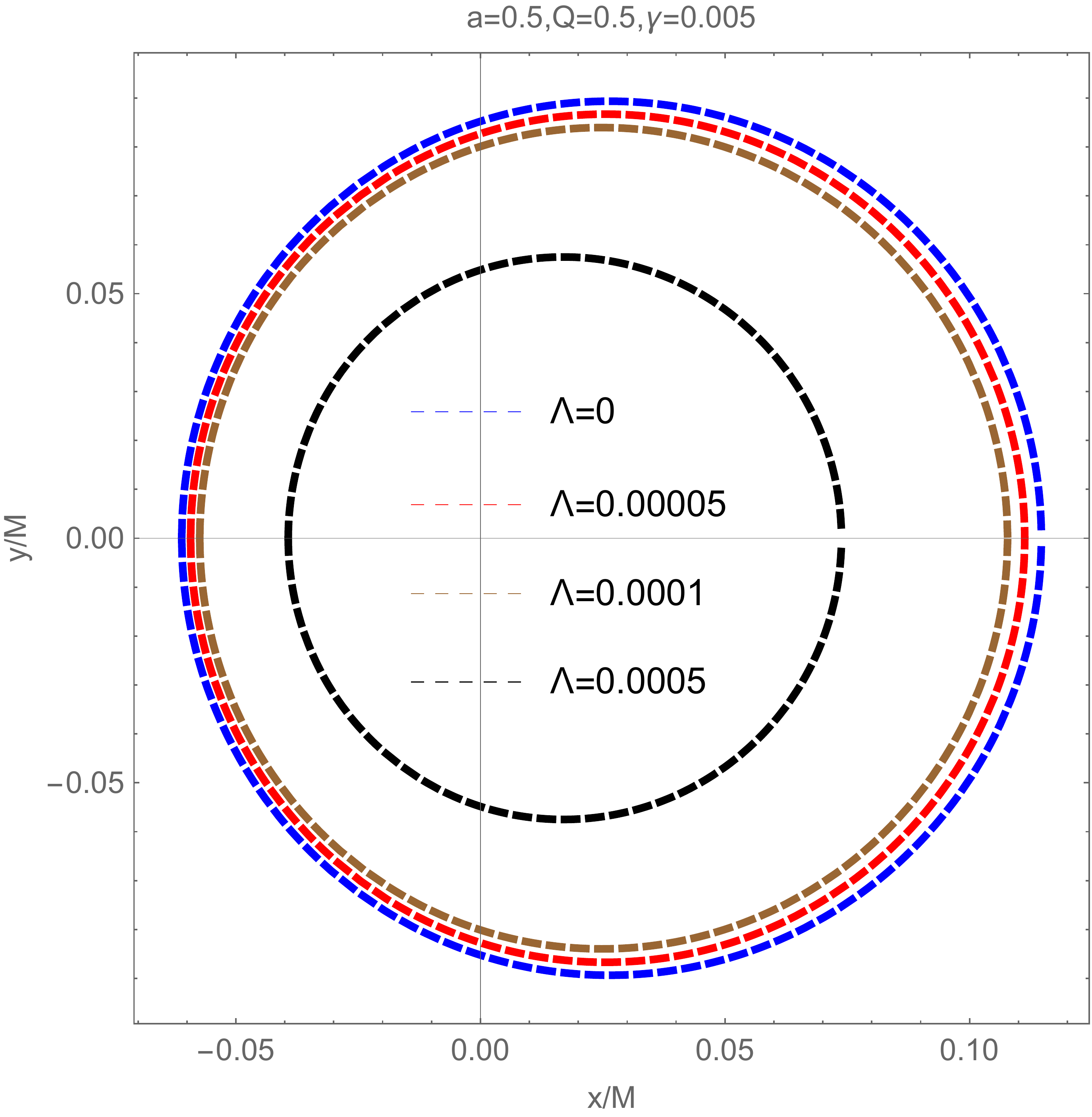}
	\includegraphics[width=.24\textwidth]{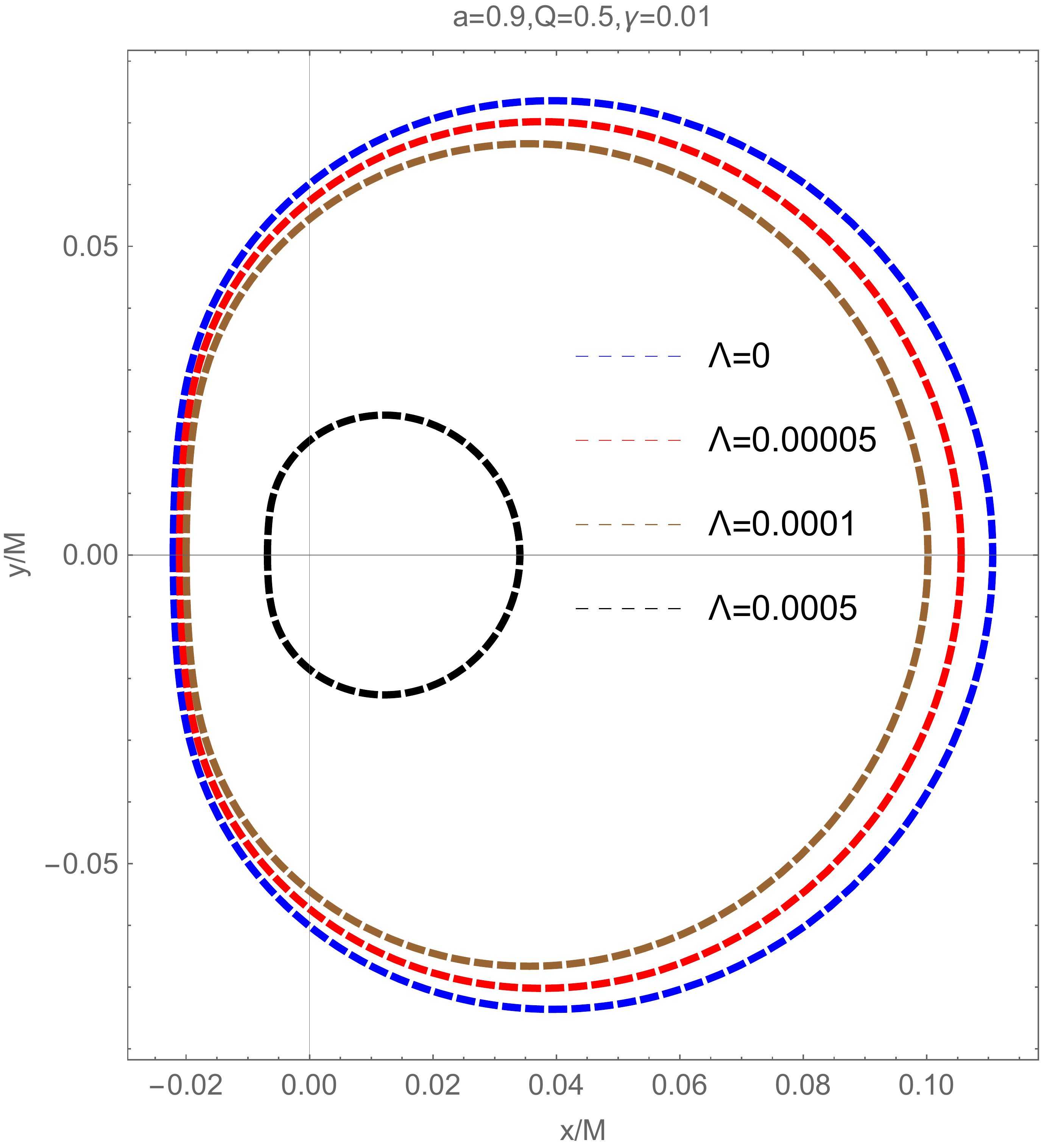}
	\caption{Shape of shadows with fixed $a$, $Q$ and $\Lambda$ for a set of values of parameters $\Lambda$}
	\label{shadow2}
\end{figure}
Fig.~\ref{shadow2} shows the effect s of cosmological constant on the shadow for different values of $a$, $Q$ and $\gamma$. From the picture, we see that the shadow maintains shapes with different values of $\Lambda$. And with a increasing values of $\Lambda$, the size of shadow decreases.

\section{Conclusions and discussions}\label{7}
  In this paper, we have obtained the  solution of a nonlinear magnetic charged rotating black hole surrounded by quintessence with a cosmological constant. The structure of the black hole horizons was studied in detail. By solving the relevant equation numerically, we found that for any fixed parameters $Q$, $\gamma$ and $\Lambda$, when $a<a_{E}$, the radii of inner horizons increase with the increasing $a$ while the radii of outer horizons decrease with $a$. At the same time, $r_{\Lambda}$ increase as $a$ increase but not so obviously. For $a=a_{E}$, we have an extremal black hole with degenerate horizons. If $a>a_{E}$, no black hole will form. Similarly, for any given values of parameters $a$, $\gamma$ and $\Lambda$, inner and outer horizons get closer first with the increase of $Q$, then coincide when $Q=Q_{E}$ and eventually disappear when $Q>Q_{E}$. And $r_{\Lambda}$ increase with increasing $Q$ but not so significantly. Furthermore,  with fixed $a$ and $Q$, the cosmological horizon $r_{\Lambda}$ significantly decrease while $r_{+}$ increase as $\gamma$ or $\Lambda$ increase. 
  
\par
 And we have discussed the behavior of photon region of in details. Through graphs, we have shown the behavior of size and shape of the photon region of our black hole by varying $a$, $Q$, $\gamma$ and $\Lambda$. And we find that the variation of photon region under $\gamma$ can be characterized into two categories.
 \par
 At last, we showed different shapes of shadow that is found by varying the intensity of quintessence, mass, magnetic charge and cosmological constant. And find the intensity of quintessence field and cosmological constant will diminish the size of shadows. 
 \par
 The existence of quintessence around black holes and cosmological constant play  important roles in many astrophysical phenomena, so our work may provide a tool for observation of quintessence. 

\section*{Conflicts of Interest}
  The authors declare that there are no conflicts of interest regarding the publication of this paper.

\section*{Acknowledgments}
  We would like to thank the National Natural Science Foundation of China (Grant No.11571342) for supporting us on this work.
  This work makes use of the Black Hole Perturbation Toolkit.

\section*{References}

 \bibliographystyle{unsrt}
 \bibliography{ref}
\end{document}